\documentclass[a4paper,11pt]{article}
\usepackage[dvipdfmx]{graphicx}
\usepackage{amsmath}
\usepackage{bm,color,url}
\usepackage{amssymb}
\usepackage{slashed}
\usepackage[top=30truemm,bottom=30truemm,left=25truemm,right=25truemm]{geometry}
\usepackage{hyperref}
\usepackage{cite}
\hypersetup{%
setpagesize=false,
 bookmarksnumbered=true,%
 bookmarksopen=true,%
 colorlinks=true,%
 linkcolor=blue,
 citecolor=red,
}

\newcommand{\ifb}{fb$^{-1}$}

\begin{document}
\titlepage


  \begin{flushright}
   {\bf
  \begin{tabular}{l}
OCHA-PP-362
  \end{tabular}
   }
  \end{flushright}

\vspace*{1.0cm}
\baselineskip 18pt
\begin{center}
 {\LARGE \bf
Probing a degenerate-scalar scenario \\
in a pseudoscalar dark-matter model
}

\vspace*{1.0cm}

 {\large \bf
 Sachiho Abe$^a$, Gi-Chol Cho$^b$ and
 Kentarou Mawatari$^c$
}

\vspace*{0.5cm}
$^a${\em Graduate school of Humanities and Sciences, Ochanomizu
 University, Tokyo 112-8610, Japan}
 \\
$^b${\em Department of Physics, Ochanomizu University, Tokyo 112-8610,
 Japan}
 \\
 $^c${\em Faculty of Education, Iwate University, Morioka, Iwate 020-8550,
  Japan}
  \\
\end{center}

\vspace*{1cm}

\baselineskip 18pt
\begin{abstract}
\noindent
We study a pseudoscalar dark-matter model
arising from a complex singlet extension of the standard model (SM), 
and show that the dark-matter--nucleon scattering is suppressed
when two CP-even scalars are degenerate.
In such a degenerate-scalar scenario we explore the model parameter space
which satisfies constraints from the direct detection experiments and the relic density
of dark matter.
In addition, we discuss a possibility to verify such a scenario
by using the recoil mass technique at the International Linear Collider. 
We find that a pair of states separated by 0.2~GeV can be distinguished
from the single SM-like Higgs state
at 5$\sigma$ with integrated luminosity of 2~ab$^{-1}$.  
\end{abstract}

\newpage
\section{Introduction}
The existence of dark matter (DM) has been suggested by several cosmological observations,
e.g. precise measurements of the cosmic microwave background~\cite{Aghanim:2018eyx}.
Although weakly interacting massive particles (WIMPs) are known as a good candidate of DM, 
no signature of the WIMP DM has been found yet in particle physics experiments.
Among various WIMP-DM searches, direct detection experiments
such as XENON1T~\cite{Aprile:2018dbl} have significantly improved the upper bounds 
on the cross section between DM and nucleon,
so that some new physics models which predict the DM candidate are severely constrained.

An interpretation of the null result in the direct DM detection experiments is 
that the DM is heavy ($\gtrsim O(\mathrm{TeV})$) or light ($\lesssim O(\mathrm{GeV})$) enough. 
Another possibility to explain the result is that effective interactions 
between DM and nucleon are suppressed by some reasons. 
For example, it has been known that 
the tree-level amplitudes of DM--quark scattering in the non-relativistic limit are proportional to the momentum transfer in pseudoscalar portal models~\cite{Ipek:2014gua,Escudero:2016gzx} 
 so that 
the amplitudes are suppressed at low-energy limit. 
The estimation of the higher-order contributions in such models 
and testability of those effects at future direct detection experiments 
are given in ref.~\cite{Abe:2018emu}.

Another example of suppression of the DM--quark scatterings at low-energy limit has also been shown
in ref.~\cite{Gross:2017dan}, where the DM is a pseudo Nambu--Goldstone boson
as a consequence of a global U(1) symmetry breaking of the scalar potential. 
Such a model is known as the minimal pseudo Nambu--Goldstone DM model
arising from a complex singlet extension of the standard model (CxSM)~\cite{Barger:2008jx}
with a softly broken U(1) symmetry by the DM mass term.  
In this model, the DM--quark scatterings are mediated by two scalar particles.  
The amplitudes of the DM--quark scattering processes mediated by each scalar particle have the same magnitude at zero momentum transfer but the opposite relative sign, 
and hence the scattering between the DM and nucleon is highly suppressed at low energy. 
The model has been studied in various contexts in, e.g., 
refs.~\cite{Azevedo:2018oxv,Azevedo:2018exj,Ishiwata:2018sdi,Huitu:2018gbc,Alanne:2018zjm,Cline:2019okt,Arina:2019tib,Glaus:2020ihj}.
It is known, however, that the minimal model suffers from so-called the domain-wall problem 
because the $Z_2$ symmetry of the scalar potential is spontaneously broken
when the singlet field develops the vacuum expectation value (VEV)~\cite{Zeldovich:1974uw}. 

In this paper, we adopt the most general renormalizable scalar potential of CxSM 
with soft breaking terms up to mass dimension two to avoid the domain-wall problem.
Since the scalar potential has more general structure than that in the minimal model,
the suppression of the DM--nucleon scatterings in the low-energy limit is no longer guaranteed. 
We show that the scattering amplitudes mediated by the two scalar particles can still be cancelled
when the masses of the two scalars are degenerate. 
In such a degenerate-scalar scenario we explore the
parameter space which satisfies the direct detection constraints and the relic density.
In addition, we discuss a possibility to verify such a scenario by using the recoil mass technique
at the International Linear Collider (ILC)~\cite{Yan:2016xyx}.
We note that the most general model of pseudoscalar DM has recently been studied 
in ref.~\cite{Alanne:2020jwx} with a focus on the gravitational-wave signal.

This paper is organized as follows.
In Sec.~\ref{model}, we introduce our DM model. 
In Sec.~\ref{supp}, we show the cancellation condition on the scattering between DM and quarks, 
and constraints on the parameter space from the direct detection experiments and 
the relic density of the DM. 
We discuss a possibility to verify a degenerate-scalar scenario at the ILC in Sec.~\ref{coll}.
Sec.~\ref{summary} is devoted to summary.

\section{Pseudoscalar dark-matter model}\label{model}

We start from the following scalar potential of the CxSM~\cite{Barger:2008jx}:
\begin{align}
 V
&=
\frac{m^2}{2}|H|^2 + \frac{\lambda}{4}|H|^4
+ \frac{\delta_2}{2}|H|^2 |S|^2 + \frac{b_2}{2}|S|^2
+ \frac{d_2}{4}|S|^4
+ \left(
 a_1 S + \frac{b_1}{4}S^2 + \mathrm{c.c.}
 \right),
\label{scalar_pot}
\end{align}
where a global U(1) symmetry for the singlet $S$ is softly broken
due to the linear ($a_1$) and quadratic ($b_1$) terms\footnote{
There are other renormalizable operators which break the global U(1) symmetry. But we have only adopted terms that close under renormalization~\cite{Barger:2008jx}. The study of ref.~\cite{Alanne:2020jwx} includes all operators which we have dropped in the scalar potential eq.~(\ref{scalar_pot}).
}. 
All the coefficients in eq.~\eqref{scalar_pot} are assumed to be real.
We note that the minimal pseudo Nambu--Goldstone DM model~\cite{Gross:2017dan}
forbids the linear term of $S$, i.e. $a_1=0$,
where the scalar potential has a $Z_2$ symmetry ($S \to -S$).
This $Z_2$ symmetry is spontaneously broken by the VEV of the singlet $S$,
and it causes the domain-wall problem~\cite{Zeldovich:1974uw}. 
The linear term ($a_1 \neq 0$) in the scalar potential, therefore, 
needs to break the $Z_2$ symmetry explicitly so that the domain-wall problem does not arise.
Although it is worth considering the UV completion of the scalar potential~\eqref{scalar_pot}, including the linear term of $S$, we do not discuss it further. 

The SM Higgs field $H$ and the singlet field $S$ are expressed in terms of the component fields as
\begin{align}
 H =
\frac{1}{\sqrt{2}}
\left(
\begin{matrix}
 0 \\
v+h
\end{matrix}
\right),
\quad
S=\frac{1}{\sqrt{2}}(v_S + s + i\chi),
\label{scalars}
\end{align}
where we adopt the unitary gauge so that the Goldstone fields in $H$ are suppressed.
The VEVs of $H$ and $S$ are denoted by $v$ and $v_S$, respectively.
The minimization conditions of the scalar potential
($\frac{\partial V}{\partial H}=0$,
 $\frac{\partial V}{\partial S}=0$)
leads to the following relations among parameters in eq.~\eqref{scalar_pot}: 
\begin{align}
 -m^2
&=
\frac{\lambda}{2}v^2 + \frac{\delta_2}{2}v_S^2,
\label{min1}
\\
-b_2
&=
\frac{\delta_2}{2}v^2
+ \frac{d_2}{2}v_S^2
+ b_1
+ 2\sqrt{2}\frac{a_1}{v_S}.
\label{min2}
\end{align}

The mass matrix $M^2$ of the CP-even scalars $(h,s)$ is given by
\begin{align}
M^2
&=
\left(
\begin{matrix}
 \frac{\lambda}{2}v^2 & \frac{\delta_2}{2} v v_S
\\
\frac{\delta_2}{2} v v_S & \Lambda^2
\end{matrix}
\right),
\end{align}
where $\Lambda^2$ is defined as
\begin{align}
\Lambda^2 &\equiv
\frac{d_2}{2}v_S^2 - \sqrt{2}\frac{a_1}{v_S}.
\end{align}
The mass eigenstates $(h_1, h_2)$ are defined by using the orthogonal matrix $O$ 
\begin{align}
 \left(
\begin{matrix}
h_1\\
h_2
\end{matrix}
\right)
=
O
 \left(
\begin{matrix}
h\\
s
\end{matrix}
\right),
\quad
O=
\left(
\begin{matrix}
 \cos\alpha & \sin\alpha
\\
-\sin\alpha & \cos\alpha
\end{matrix}
\right).
\label{omatrix}
\end{align}
Then, the mass eigenvalues ($m_{h_1},m_{h_2}$) and the mixing angle $\alpha$
are given by
\begin{align}
m^2_{h_1,h_2}
 &=
\frac{1}{2}
\left\{
\frac{\lambda}{2}v^2
+ \Lambda^2
\mp
\sqrt{
\left(\frac{\lambda}{2}v^2 - \Lambda^2\right)^2
+
\left(\delta_2v v_S
\right)^2
}
\right\},
\\
 \cos 2\alpha
&= \frac{
\frac{\lambda}{2}v^2 - \Lambda^2
}{m_{h_1}^2-m_{h_2}^2}.
\end{align}
Since the global U(1) symmetry is softly broken,
the CP-odd scalar $\chi$ has a mass $m_\chi$ as
\begin{align}
 m_\chi^2 =  -b_1 - \sqrt{2} \frac{a_1}{v_S}.
\end{align}
The CP symmetry of the scalar potential~\eqref{scalar_pot} forbids
the pseudoscalar $\chi$ to decay into the CP-even scalars
so that $\chi$ could be identified as a DM particle.

We give the interaction Lagrangians which are necessary to our study below.
The scalar-trilinear interactions for $\chi$ and $h_1$ or $h_2$ are given by
\begin{align}
  \mathcal{L}_S
 &=  -\frac{1}{2v_S}
 \left\{
 \left(
   m_{h_1}^2 +\frac{\sqrt{2}a_1}{v_S}
 \right)\sin\alpha\,
 h_1 \chi^2
 +
 \left(
   m_{h_2}^2 + \frac{\sqrt{2}a_1}{v_S}
 \right)\cos\alpha\,
 h_2 \chi^2
 \right\}.
\label{trilinear}
\end{align}
The Yukawa interactions of a fermion $f$ and the CP-even scalar $h_1$ or $ h_2$ 
are given by
\begin{align}
 \mathcal{L}_Y
 &=
 -\frac{m_f}{v} \bar{f} f \left(
 h_1 \cos\alpha - h_2 \sin\alpha
 \right),
\end{align}
where $m_f$ denotes the mass of the fermion $f$.

Here, we summarize the theoretical constraints on the model parameters.
In the scalar potential, the quartic couplings $\lambda$ and $d_2$ are required
to satisfy the following conditions from the perturbative unitarity~\cite{Chen:2014ask}
\begin{align}
 \lambda < \frac{16\pi}{3},\quad d_2 < \frac{16\pi}{3}.
 \label{th_per}
\end{align}
Also the stability condition of the scalar potential is given by~\cite{Barger:2008jx}
\begin{align}
 \lambda
\left(
d_2 - \frac{2\sqrt{2}a_1}{v_S^3}
\right) > \delta_2^2.
\label{th_stab}
\end{align}

We finally summarize the model parameters which will be used in the following analyses. 
There are seven parameters in the scalar potential and two VEVs.
Since there are two minimization conditions on the scalar potential
in eqs.~\eqref{min1} and \eqref{min2}, 
the number of the model parameters is seven.
We take $v\simeq246$~GeV and identify $h_1$ as a scalar particle
which has been discovered at the LHC with $m_{h_1}\simeq125$~GeV.
We choose the remaining five parameters
\begin{align}
   m_{h_2},\ m_\chi,\ \alpha,\ a_1,\ v_S
\label{inputs}
\end{align}
as inputs so that other parameters
\begin{align}
m^2,\ \lambda,\ \delta_2,\ b_2,\ d_2,\ b_1
\label{outputs}
\end{align}
are given as functions of~\eqref{inputs}. 
It is convenient to express output parameters~\eqref{outputs}
by inputs~\eqref{inputs} for later discussions; 
\begin{align}
 m^2 
&=
 -\frac{\lambda}{2}v^2 - \frac{\delta_2}{2}v_S^2, 
\\	
 \lambda
&=
 \frac{2}{v^2}\left(
 m_{h_1}^2 \cos^2\alpha 
 +m_{h_2}^2 \sin^2\alpha 
\right), 
\\
 \delta_2 
 &=
\frac{\frac{\lambda}{2}v^2- \Lambda^2}{v v_S}
\tan 2\alpha, 
\\
 b_2 
&=
 -\frac{\delta_2}{2}v^2 - \frac{d_2}{2}v_S^2 - b_1 
 - 2\sqrt{2} \frac{a_1}{v_S},
 \\
 d_2 
&=
 2\left(\frac{m_{h_1}^2}{v_S}\right)^2 \sin^2\alpha 
 + 2\left(\frac{m_{h_2}^2}{v_S}\right)^2 \cos^2\alpha 
 + 2\sqrt{2}\frac{a_1}{v_S^3}, 
\\
 b_1 
 &=
 -m_\chi^2 - \sqrt{2}\frac{a_1}{v_S}.
\end{align}

\section{Suppression of dark-matter--nucleon scattering}\label{supp}

\begin{figure} 
 \center
 \includegraphics[width=3cm]{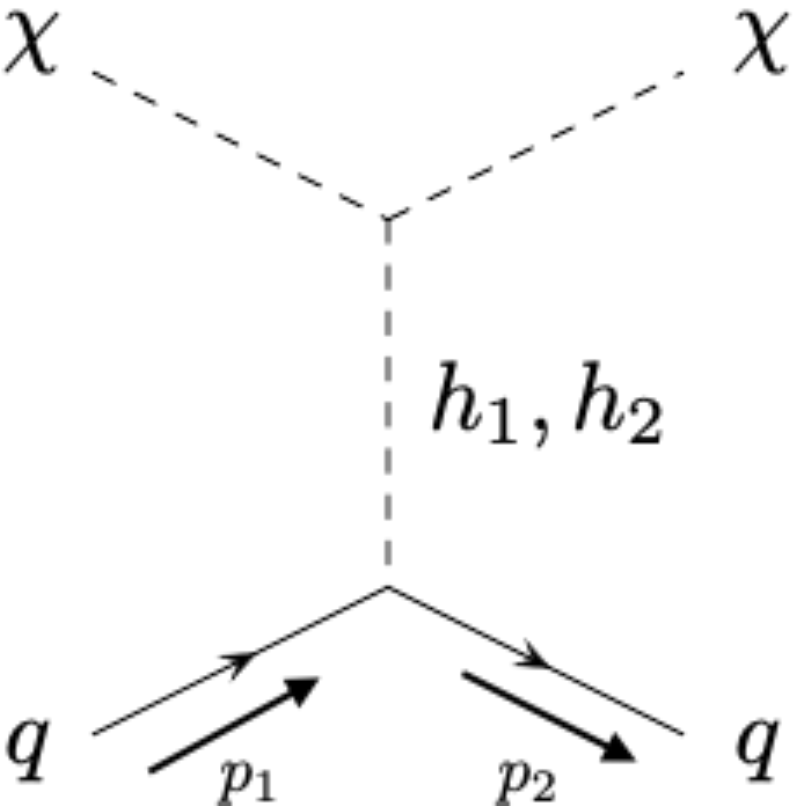}
 \caption{
  Feynman diagram of the scattering process $\chi q \to \chi q$
  mediated by CP-even scalars $h_1$ and $h_2$.
 }
 \label{fig:feyn_diag}
\end{figure}

In this section, we discuss a suppression mechanism
of the scattering process of the DM $\chi$ off a quark $q$,
$\chi q \to \chi q$,
whose Feynman diagram is shown in Fig.~\ref{fig:feyn_diag}. 
We also discuss constraints on the model parameter space
from the direct detection experiments and the relic density of the DM. 

The amplitudes mediated by $h_1$ and $h_2$ are given by
\begin{align}
  i\mathcal{M}_{h_1}
 &=
 -i \frac{m_q}{vv_S}
 \frac{
 m_{h_1}^2 + \frac{\sqrt{2}a_1}{v_S}
 }{t-m_{h_1}^2}
 \sin\alpha \cos\alpha\,
 \bar{u}(p_2) u(p_1),
 \label{amp1}
 \\
  i\mathcal{M}_{h_2}
 &=
 +i \frac{m_q}{vv_S}
 \frac{
 m_{h_2}^2 + \frac{\sqrt{2}a_1}{v_S}
 }{t-m_{h_2}^2}
 \sin\alpha \cos\alpha\,
 \bar{u}(p_2) u(p_1),
\label{amp2}
\end{align}
where $t \equiv (p_1-p_2)^2$, and $p_1$ ($p_2$)
is the momentum of quark $q$ in the initial (final) state.
Since the momentum transfer $t$ in the process is very small
with respect to the mass of the scalars $m_{h_{1,2}}$,
the sum of the two amplitudes can be written as
\begin{align}
   i(\mathcal{M}_{h_1}+\mathcal{M}_{h_2})
 &\simeq
 i\frac{m_q}{v v_S}  \sin\alpha\cos\alpha\,\bar{u}(p_2) u(p_1)
 \nonumber \\
 &
 \times\frac{1}{m_{h_1}^2m_{h_2}^2}
 \left\{
 \left(
 m_{h_2}^2-m_{h_1}^2
 +\frac{\sqrt{2}a_1}{v_S}
  \frac{m_{h_2}^4-m_{h_1}^4}{m_{h_1}^2m_{h_2}^2}
 \right)t
 +
 \frac{\sqrt{2}a_1}{v_S}
 \left(
  m_{h_2}^2-m_{h_1}^2
 \right)
 \right\}.
 \label{sumamp}
\end{align}
The first term in the curly braces in r.h.s is negligible
when $t\simeq0$ as shown in ref.~\cite{Gross:2017dan}.
The second term, on the other hand, vanishes
when the two scalars are degenerate ($m_{h_1} = m_{h_2}$).
The cancellation of the two amplitudes in the degenerate limit is 
virtue of the orthogonal condition of the matrix $O$ in eq.~(\ref{omatrix}),
$O_{ik} O_{jk} = \delta_{ij}$.
Constraints on this model from the direct detection experiments,
therefore, are weaken
when the mass of the CP-even scalar $h_2$ is close to the mass of $h_1$,
$m_{h_1}\simeq 125$~GeV.
This motivates us to study phenomenological consequences
of such a degenerate-scalar scenario in the pseudoscalar DM model.
We note that in the $m_{h_1}\approx m_{h_2}$ case 
the two amplitudes in eqs.~\eqref{amp1} and \eqref{amp2} are cancel
even not for low energy.
This fact is relevant to understand the DM relic density in this model,
which will be discussed later.

Before turning into numerical studies, we note that
the scattering between DM and quarks at 1-loop level
in the minimal model of \cite{Gross:2017dan}
has been studied in refs.~\cite{Azevedo:2018exj,Ishiwata:2018sdi,Glaus:2020ihj}.
Although the cancellation of the amplitudes at the low-energy limit
does not hold any more at 1-loop level,
the contributions are small enough to be neglected.  
Especially, for the $m_{h_1}\approx m_{h_2}$ case,
those contributions are strongly suppressed~\cite{Azevedo:2018exj},
which can be applied for our degenerate scenario even at 1-loop level.
  
In the following numerical studies, we implemented the pseudoscalar DM model (i.e. the CxSM)
as the {\sc UFO} format~\cite{Degrande:2011ua}
by using {\sc FeynRules v2.3}~\cite{Alloul:2013bka},
and used it in {\sc MadDM~v3.0}~\cite{Backovic:2013dpa,Backovic:2015cra,Ambrogi:2018jqj}
to compute the DM--nucleon scattering cross section and the relic density of DM. 

\begin{figure}
 \center
 \includegraphics[width=0.325\textwidth]{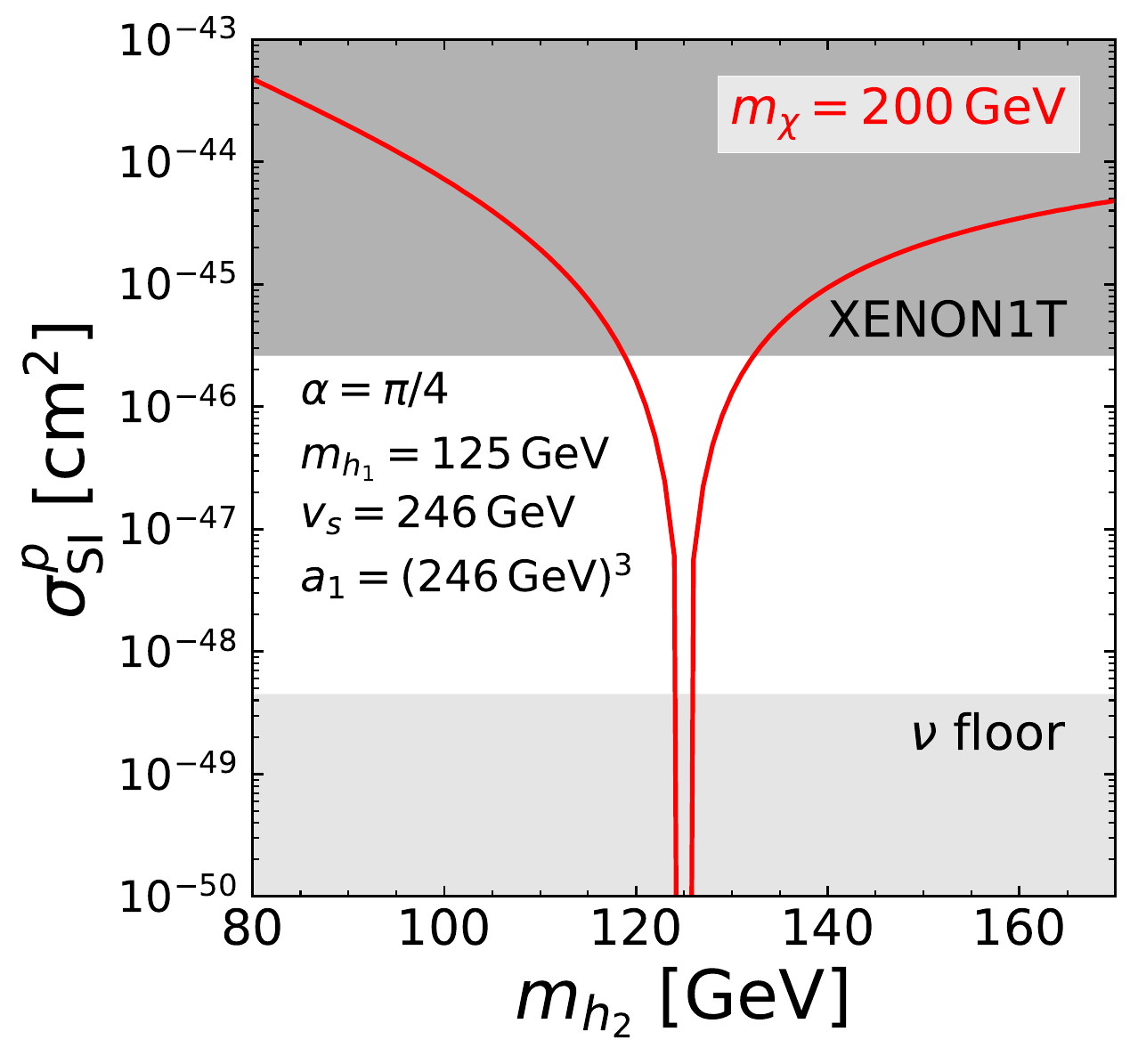} 
 \includegraphics[width=0.325\textwidth]{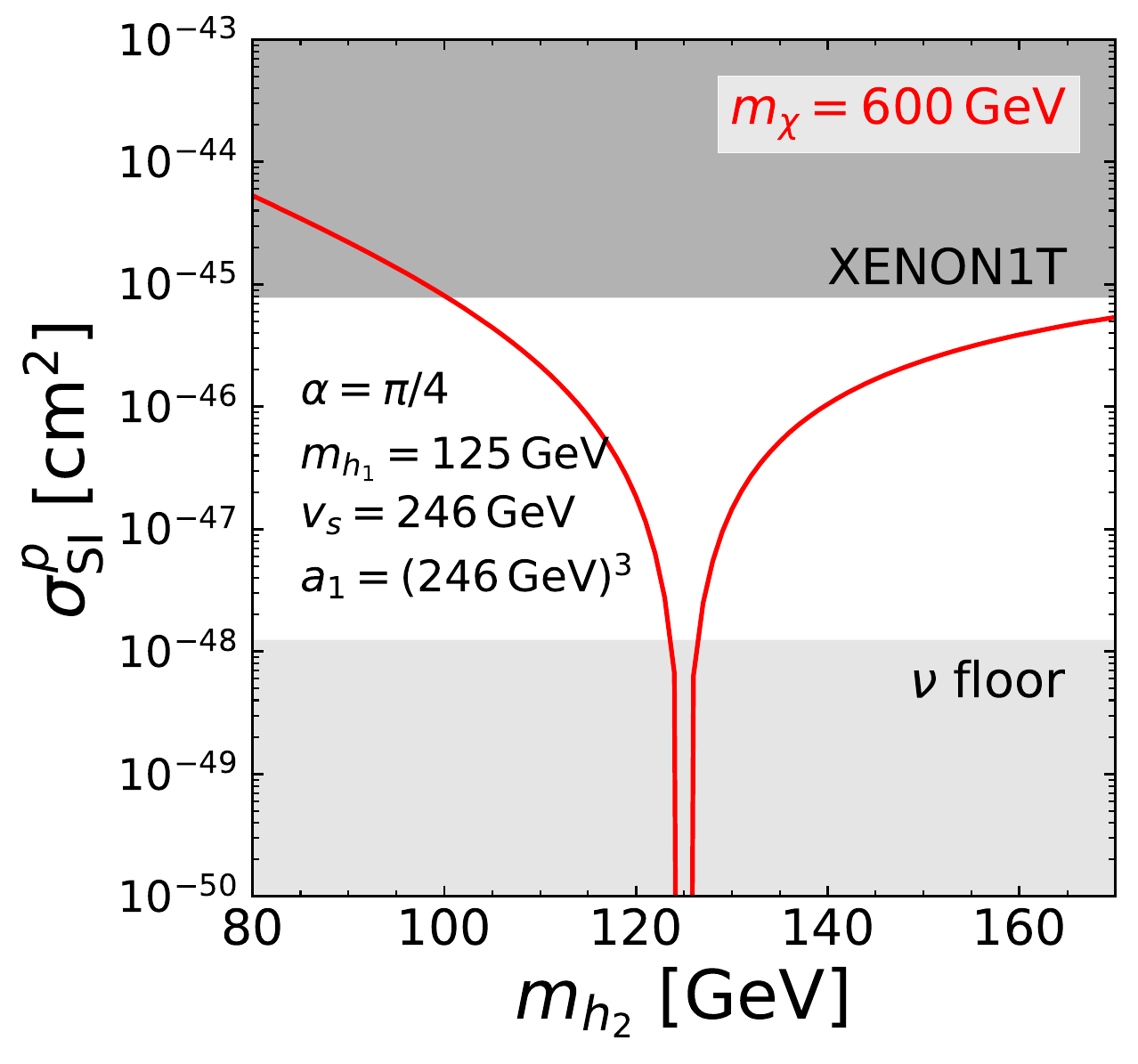}
 \includegraphics[width=0.325\textwidth]{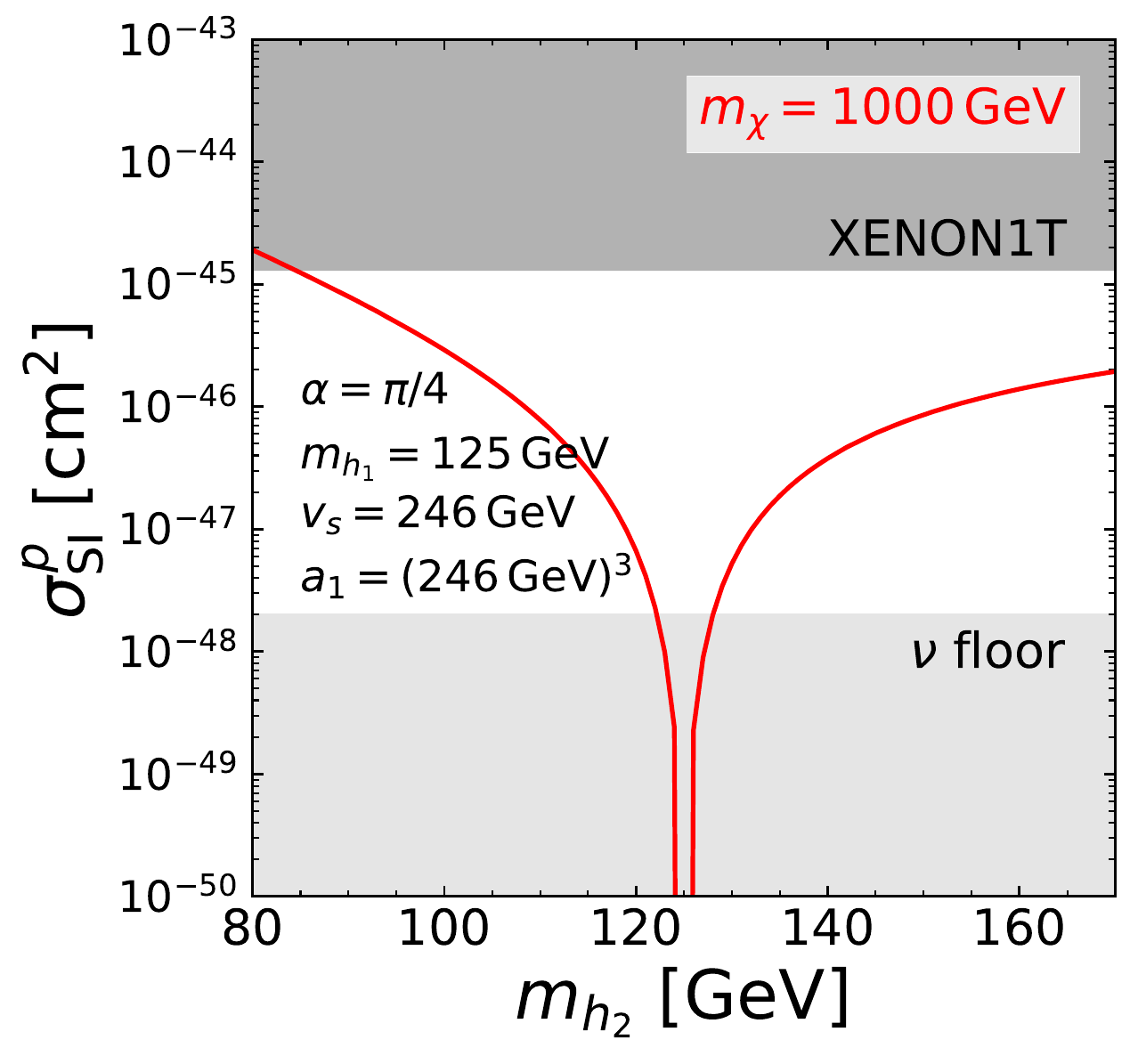}
 \caption{
  Spin-independent DM--nucleon scattering cross section as a function of $m_{h_2}$
  for the DM mass fixed at 200, 600 and 1000~GeV from the left to right panels, respectively.
 }
 \label{fig:fig_DM_mh}
\end{figure}

In Fig.~\ref{fig:fig_DM_mh}, we show the spin-independent cross section
for the scattering between the DM and a nucleon $\sigma^p_{\mathrm{SI}}$ as a function of $m_{h_2}$.
Three figures correspond to the DM mass $m_\chi=200~\mathrm{GeV}$ (left),
$600~\mathrm{GeV}$ (center) and $1000~\mathrm{GeV}$ (right), respectively.
Other parameters are fixed at $\alpha=\pi/4$, $v_S=\sqrt[3]{a_1}=246~\mathrm{GeV}$ as an example.
The upper and lower shaded regions represent the excluded region
by the XENON1T experiment~\cite{Aprile:2018dbl} 
and the background from the elastic neutrino-nucleus scattering
(so-called neutrino floor)~\cite{Billard:2013qya}.
As $a_1\ne0$ in our model, the larger the mass difference between $h_1$ and $h_2$ is,
the larger the DM--nucleon scattering cross section is.
As expected, the cross section is highly suppressed around $m_{h_2}\sim m_{h_1}$. 

We now examine dependences of the DM--nucleon cross section on the other parameters.
In the following analyses we set the mixing angle $\alpha=\pi/4$, 
where the cross section is maximal. 
Then, the remaining four parameters in \eqref{inputs};
the mass difference $\Delta m=m_{h_2}-m_{h_1}$ (instead of $m_{h_2}$), 
the DM mass $m_\chi$, the tad-pole coupling $a_1$ and the VEV of the singlet field $v_S$,
are constrained from the experimental data.  

\begin{figure}
    \begin{tabular}{ccc}
      \begin{minipage}[t]{0.3\hsize}
        \centering
        \includegraphics[keepaspectratio, scale=0.33]{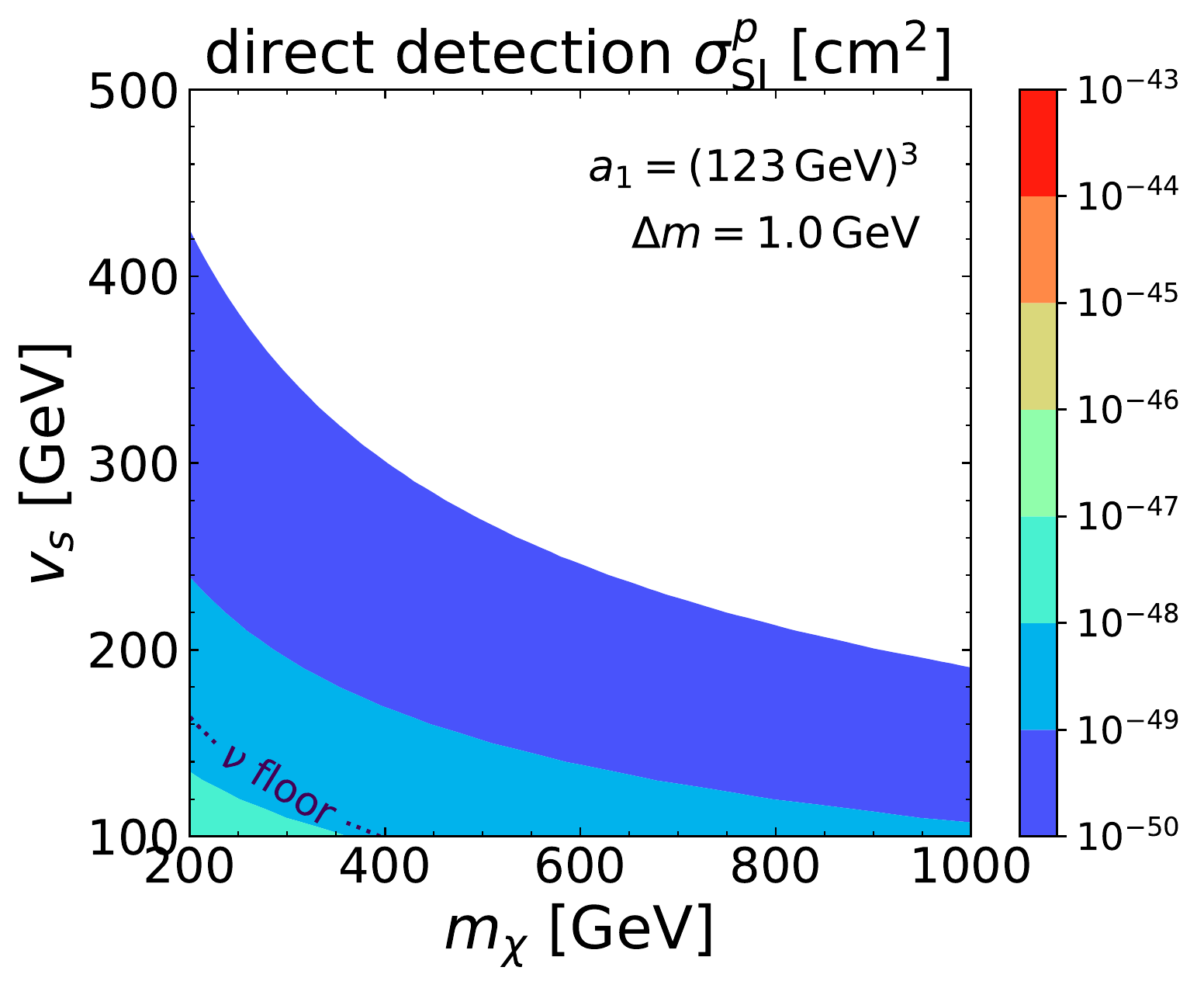}
      \end{minipage} &
      \begin{minipage}[t]{0.3\hsize}
        \centering
        \includegraphics[keepaspectratio, scale=0.33]{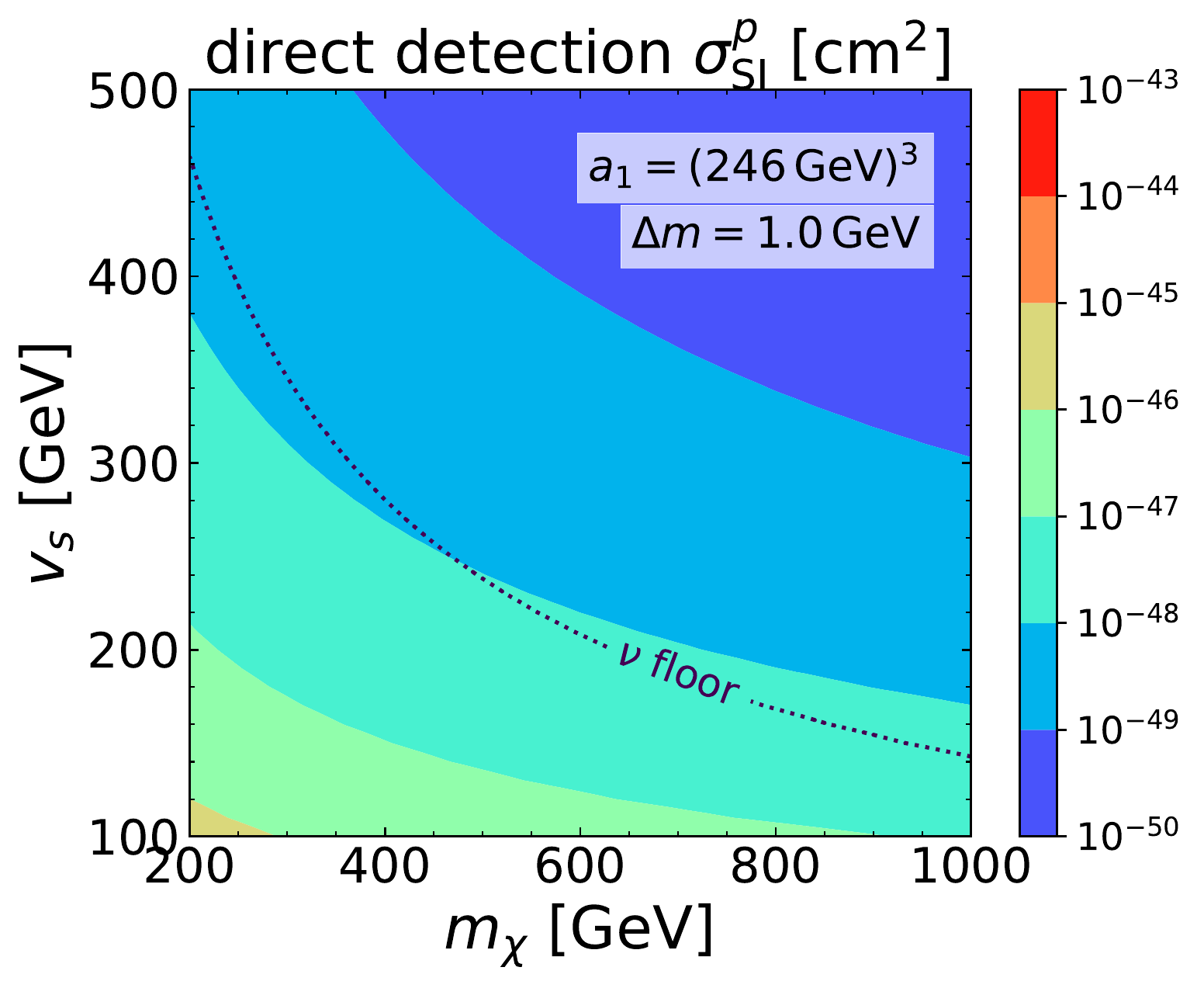}
      \end{minipage} &
      \begin{minipage}[t]{0.3\hsize}
        \includegraphics[keepaspectratio, scale=0.33]{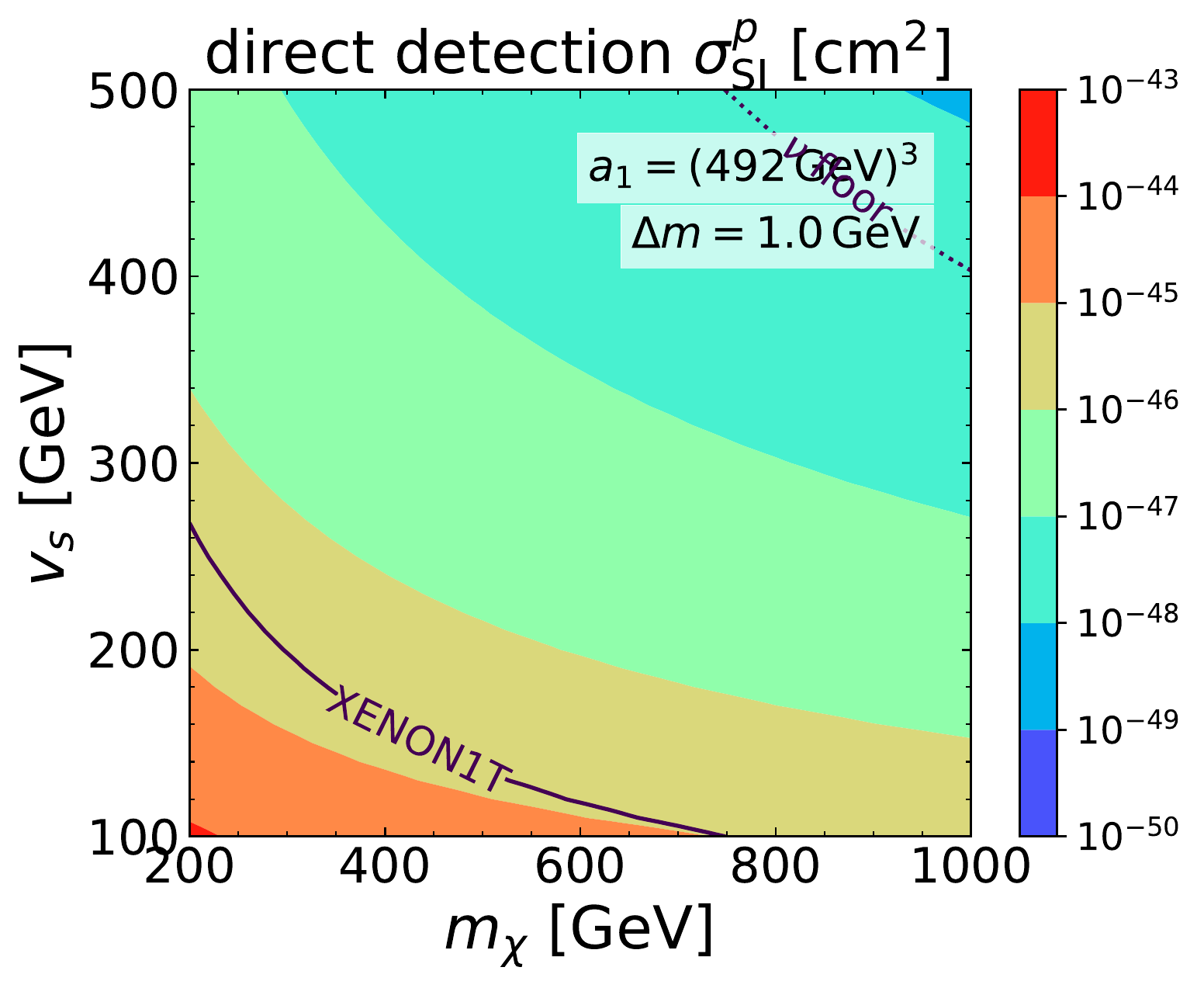}
      \end{minipage}
\\
      \begin{minipage}[t]{0.3\hsize}
        \centering
        \includegraphics[keepaspectratio, scale=0.33]{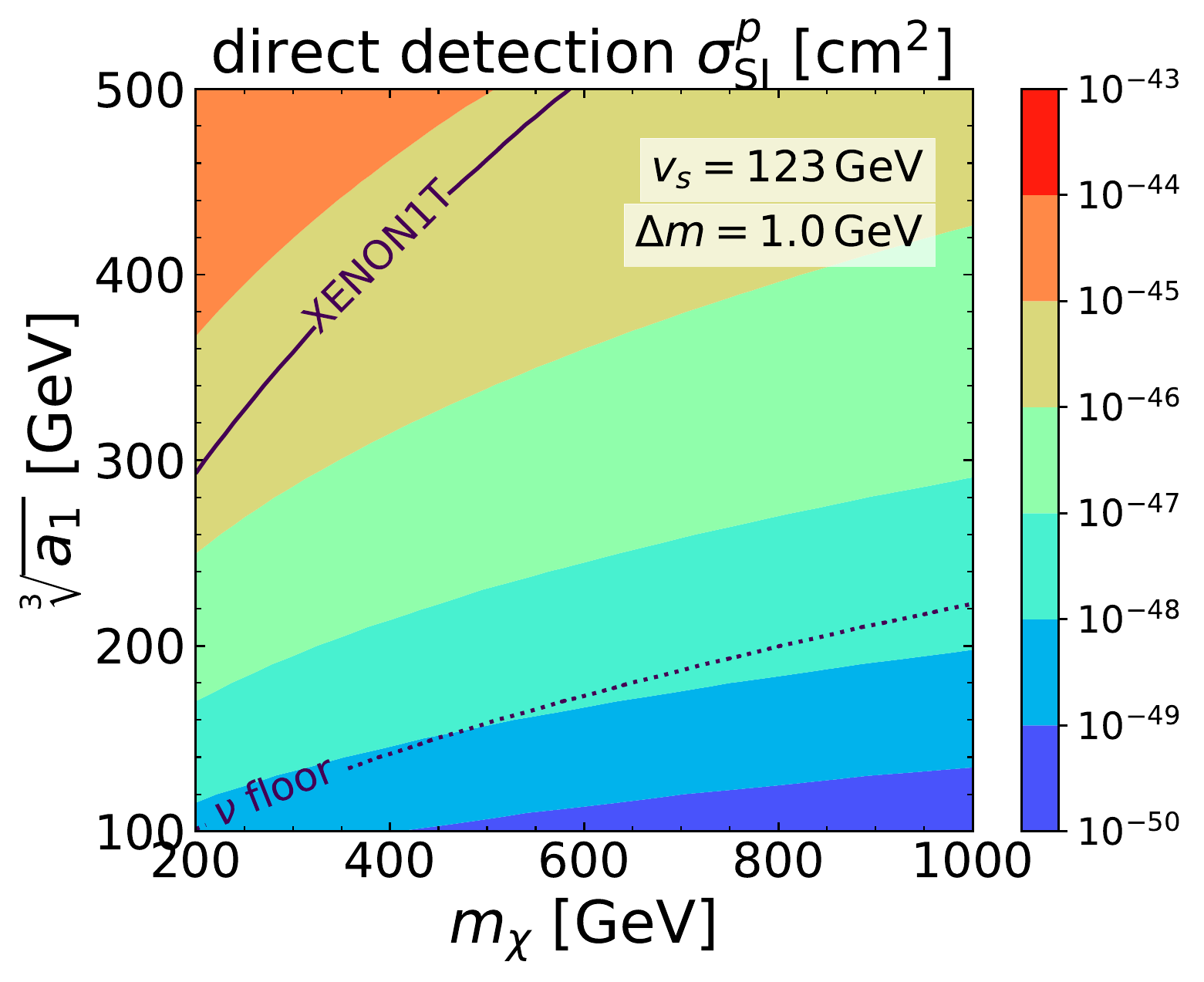}
      \end{minipage} &
      \begin{minipage}[t]{0.3\hsize}
        \centering
        \includegraphics[keepaspectratio, scale=0.33]{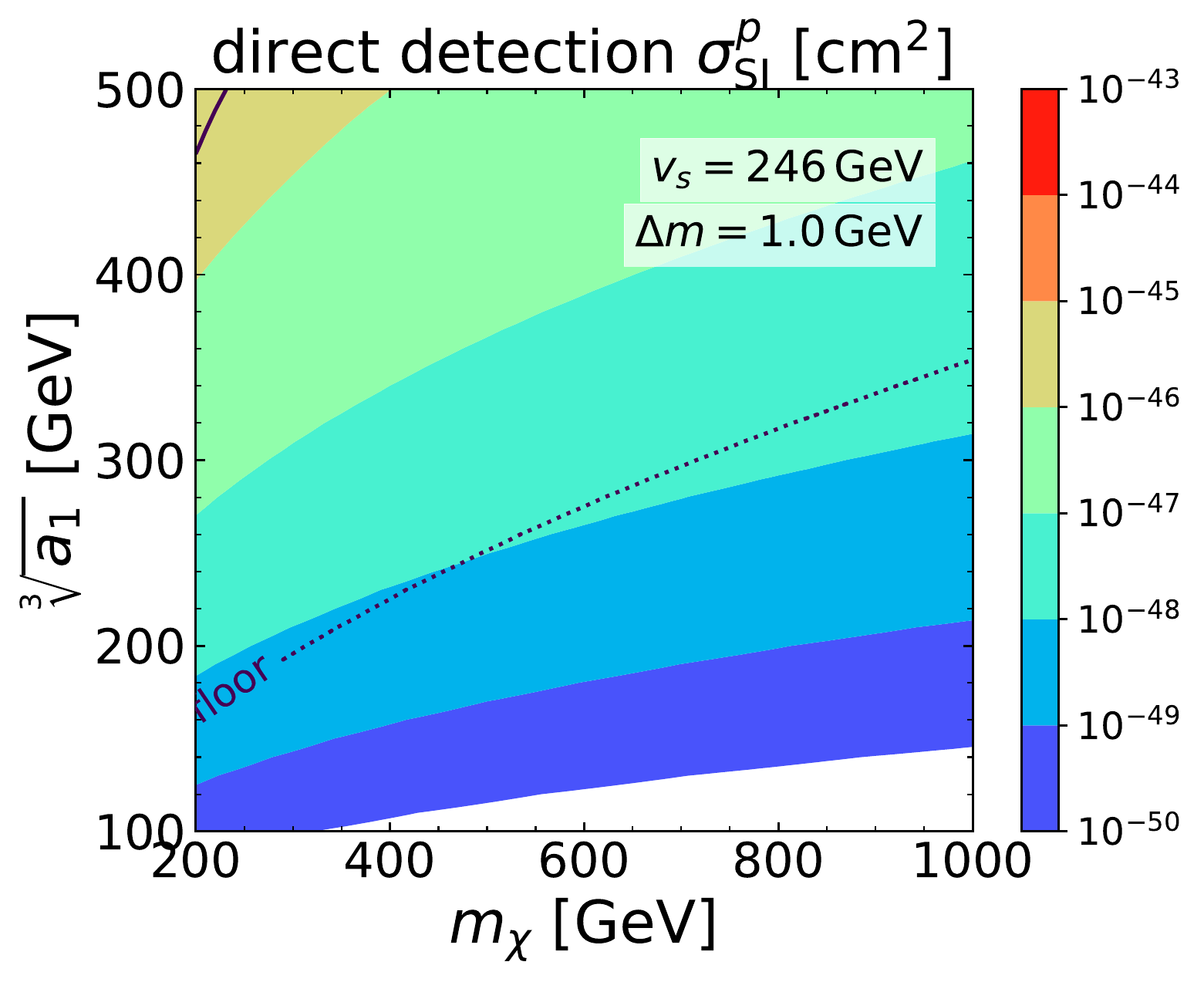}
      \end{minipage} &
      \begin{minipage}[t]{0.3\hsize}
        \includegraphics[keepaspectratio, scale=0.33]{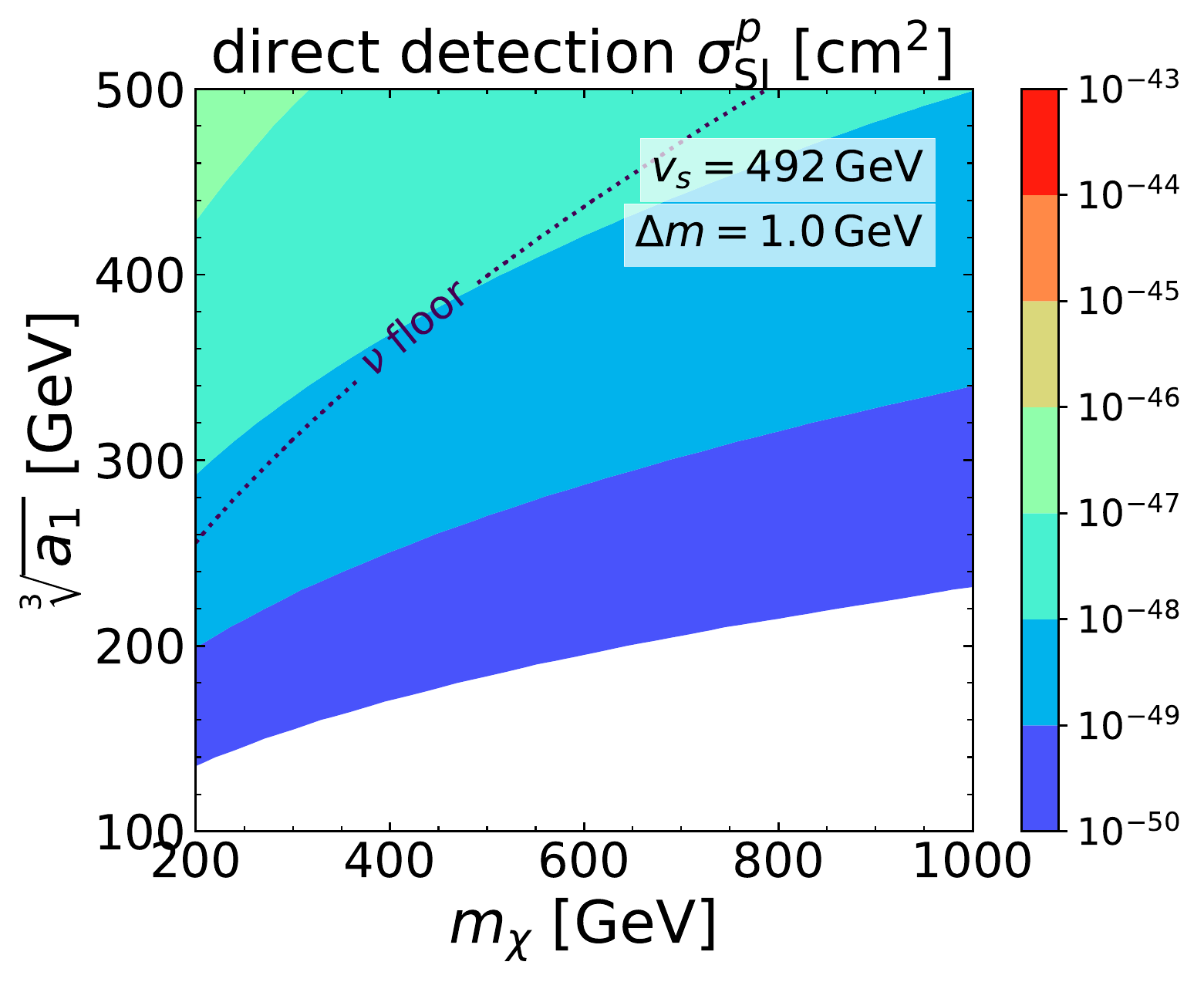}
      \end{minipage}
\\
      \begin{minipage}[t]{0.3\hsize}
        \centering
        \includegraphics[keepaspectratio, scale=0.33]{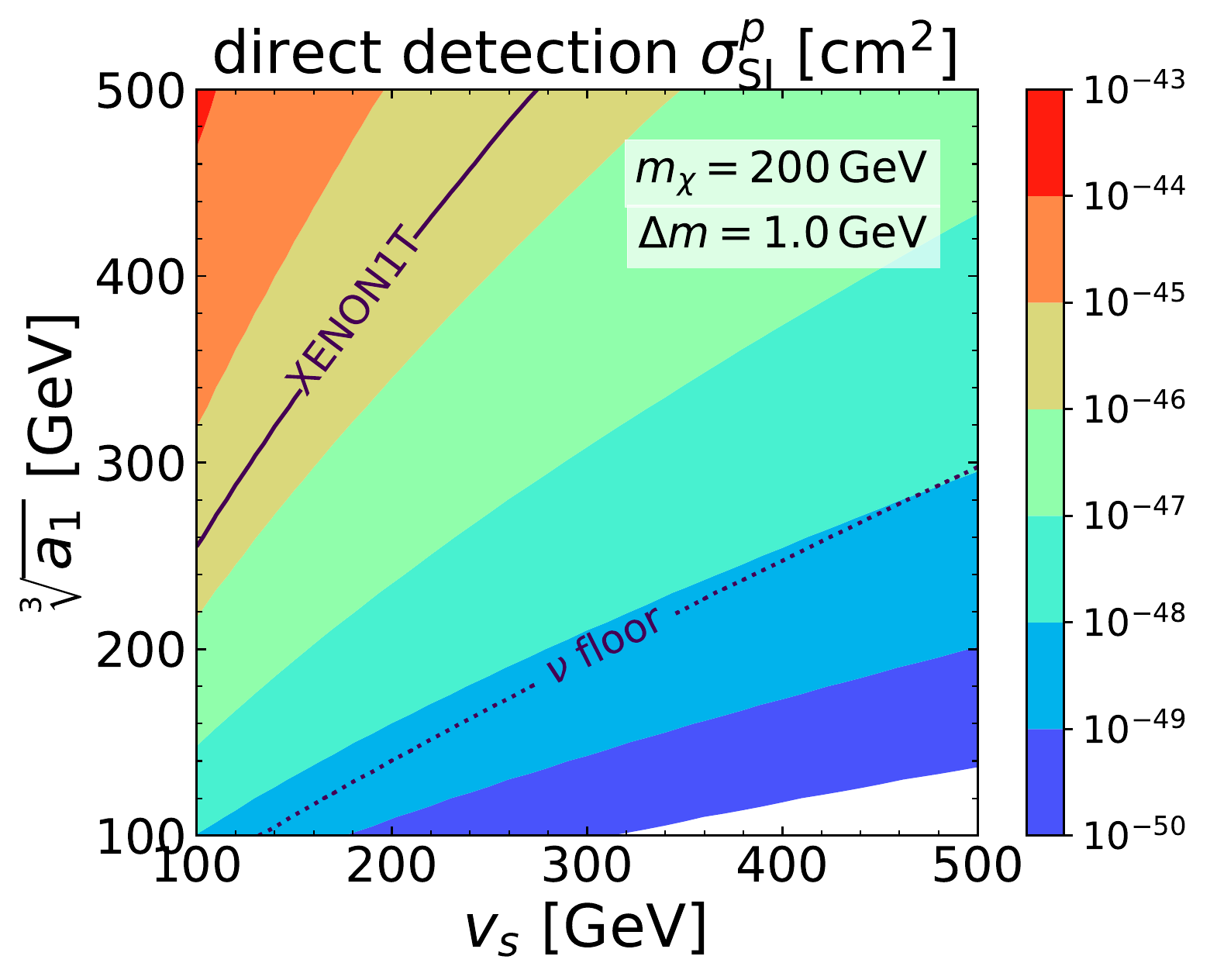}
      \end{minipage} &
      \begin{minipage}[t]{0.3\hsize}
        \centering
        \includegraphics[keepaspectratio, scale=0.33]{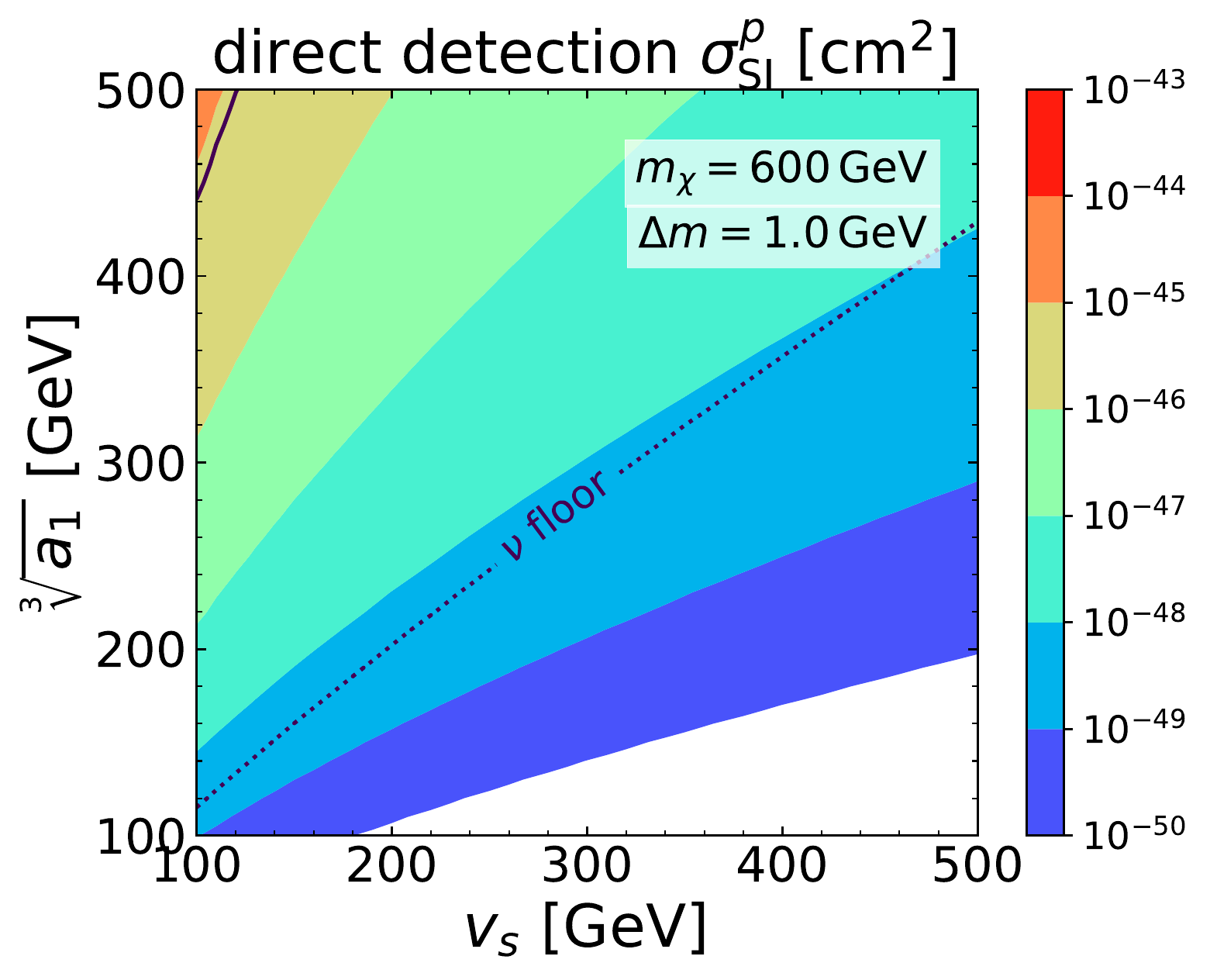}
      \end{minipage} &
      \begin{minipage}[t]{0.3\hsize}
        \includegraphics[keepaspectratio, scale=0.33]{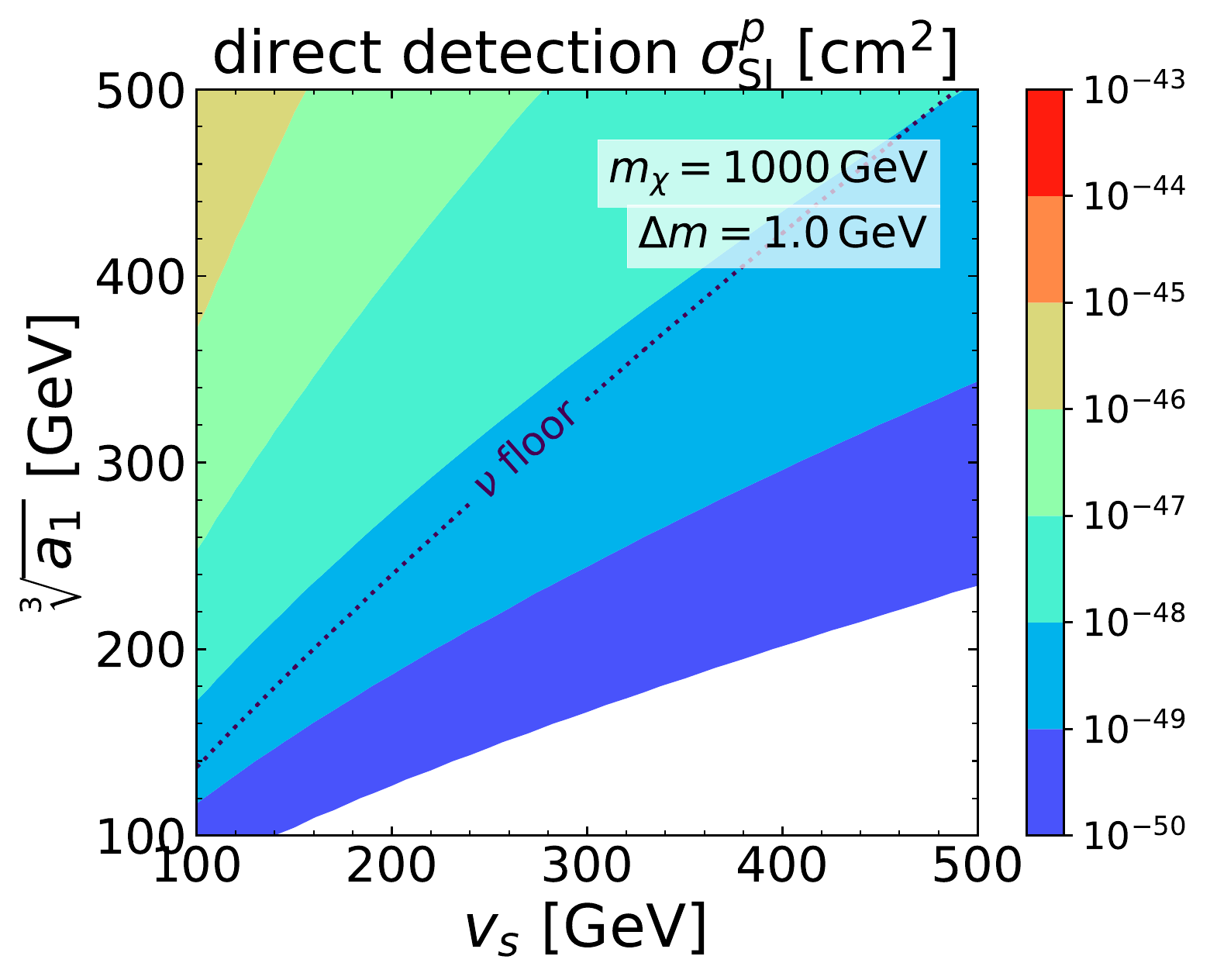}
      \end{minipage}
    \end{tabular}
 \caption{
 Spin-independent DM--nucleon scattering cross section on the $m_\chi$--$v_s$ (top),
 $m_\chi$--$\sqrt[3]{a_1}$ (middle), and $v_s$--$\sqrt[3]{a_1}$ (bottom) planes.
 $\Delta m=m_{h_2}-m_{h_1}$ is fixed at 1~GeV for all the panels, 
 while the other parameter, $a_1$ (top), $v_s$ (middle), and $m_\chi$ (bottom), 
 is taken as the low to high scale from the left to right panels.  
}
\label{fig:dd}
\end{figure}

In Fig.~\ref{fig:dd},  
we show contour plots of the spin-independent cross section $\sigma^p_{\mathrm{SI}}$ 
on $m_\chi$--$v_S$ (top), $m_\chi$--$\sqrt[3]{a_1}$ (middle) and 
$v_s$--$\sqrt[3]{a_1}$ (bottom) planes. 
The upper limit on $\sigma^p_{\mathrm{SI}}$ from XENON1T~\cite{Aprile:2018dbl} is shown
by the black-solid curve 
while the dotted-curve denotes the neutrino floor~\cite{Billard:2013qya}. 
The mass difference $\Delta m$ is fixed at $1~\mathrm{GeV}$ for all the panels. 

Three graphs in the first row of Fig.~\ref{fig:dd} correspond to 
$a_1=(123~\mathrm{GeV})^3, (246~\mathrm{GeV})^3$, and $(492~\mathrm{GeV})^3$, respectively. 
In these graphs, the cross section $\sigma^p_{\mathrm{SI}}$ decreases for larger $v_S$. 
The reason can be understood as follows. 
For larger $v_S$, the $\chi$-$\chi$-$h_i$ interactions are relatively dominated by the $\chi$-$\chi$-$s$ interaction, because it is proportional to $v_S$ while the $\chi$-$\chi$-$h$ interaction is proportional to $v$.%
\footnote{Recall that $\chi$-$\chi$-$s$ and $\chi$-$\chi$-$h$ interactions are given by $|S|^4$ and $|H|^2 |S|^2$ terms in \eqref{scalar_pot}, respectively. }
Since the singlet $s$ does not interact with quarks, the scattering processes between the DM and quarks are suppressed for larger $v_S$. 
This is why the scattering amplitude of the DM and quarks in eq.~\eqref{sumamp} 
is inversely proportional to $v_S$.
On the other hand, 
the cross section increases as $a_1$ increases from the left to right panels.
This is because the scattering amplitude of eq.~(\ref{sumamp}) is proportional to $a_1$. 
As seen, for the large $a_1$ case, the small $m_\chi$ and $v_s$ region is already excluded
by the current direct detection experiment.
Otherwise, the whole parameter space on this plane satisfies the constraint.

\begin{figure}
    \begin{tabular}{ccc}
      \begin{minipage}[t]{0.3\hsize}
        \centering
        \includegraphics[keepaspectratio, scale=0.33]{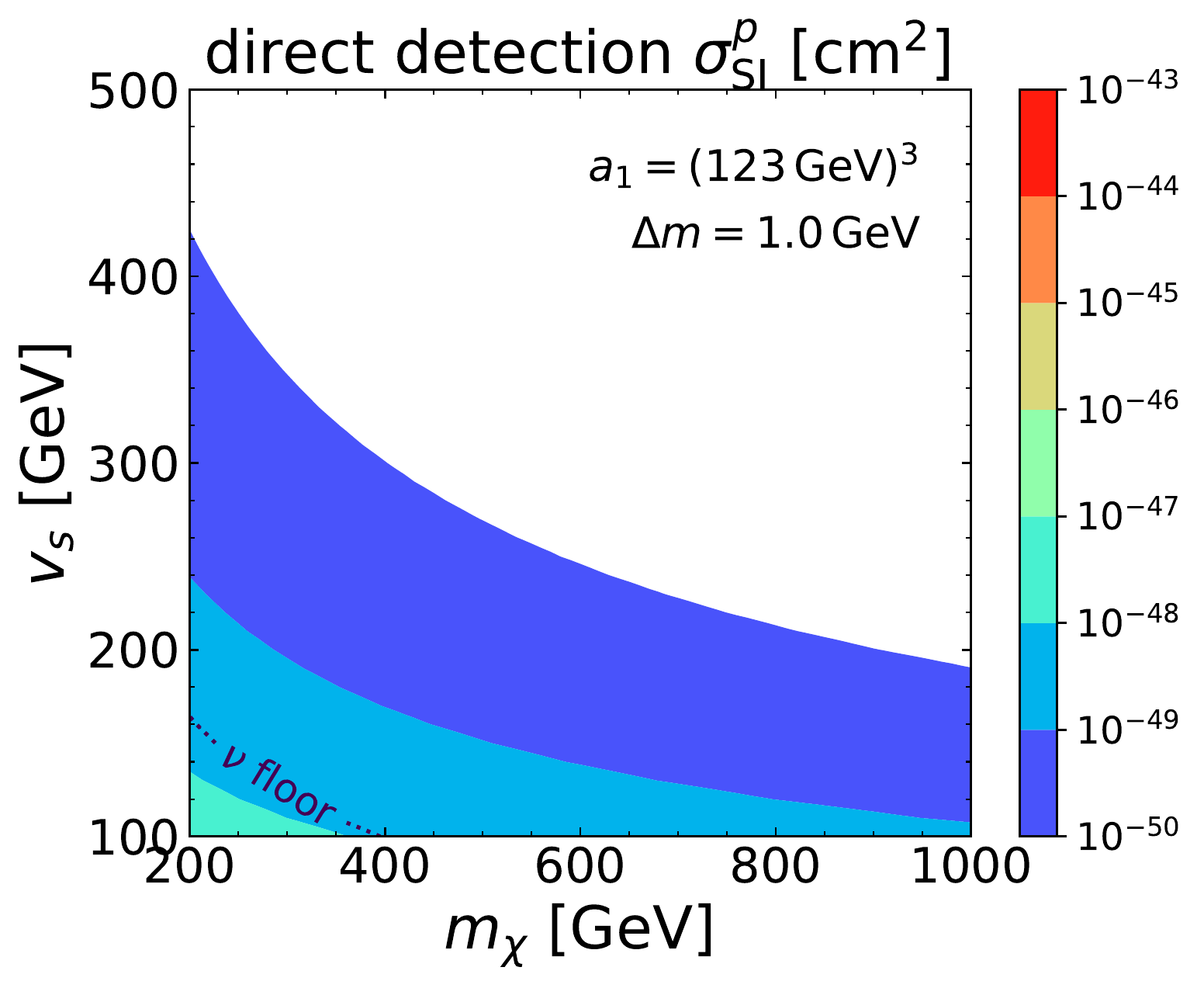}
      \end{minipage} &
      \begin{minipage}[t]{0.3\hsize}
        \centering
        \includegraphics[keepaspectratio, scale=0.33]{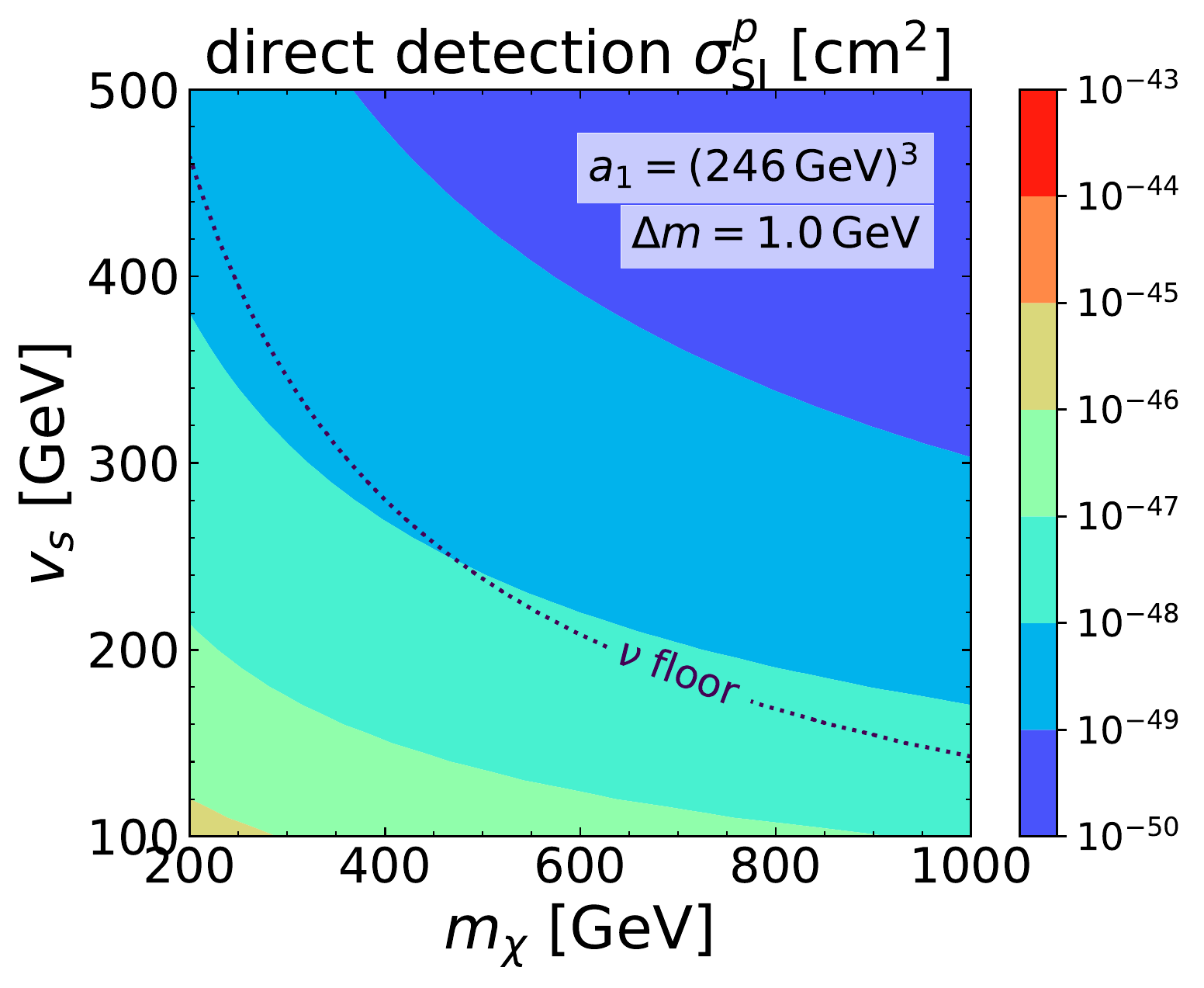}
      \end{minipage} &
      \begin{minipage}[t]{0.3\hsize}
        \includegraphics[keepaspectratio, scale=0.33]{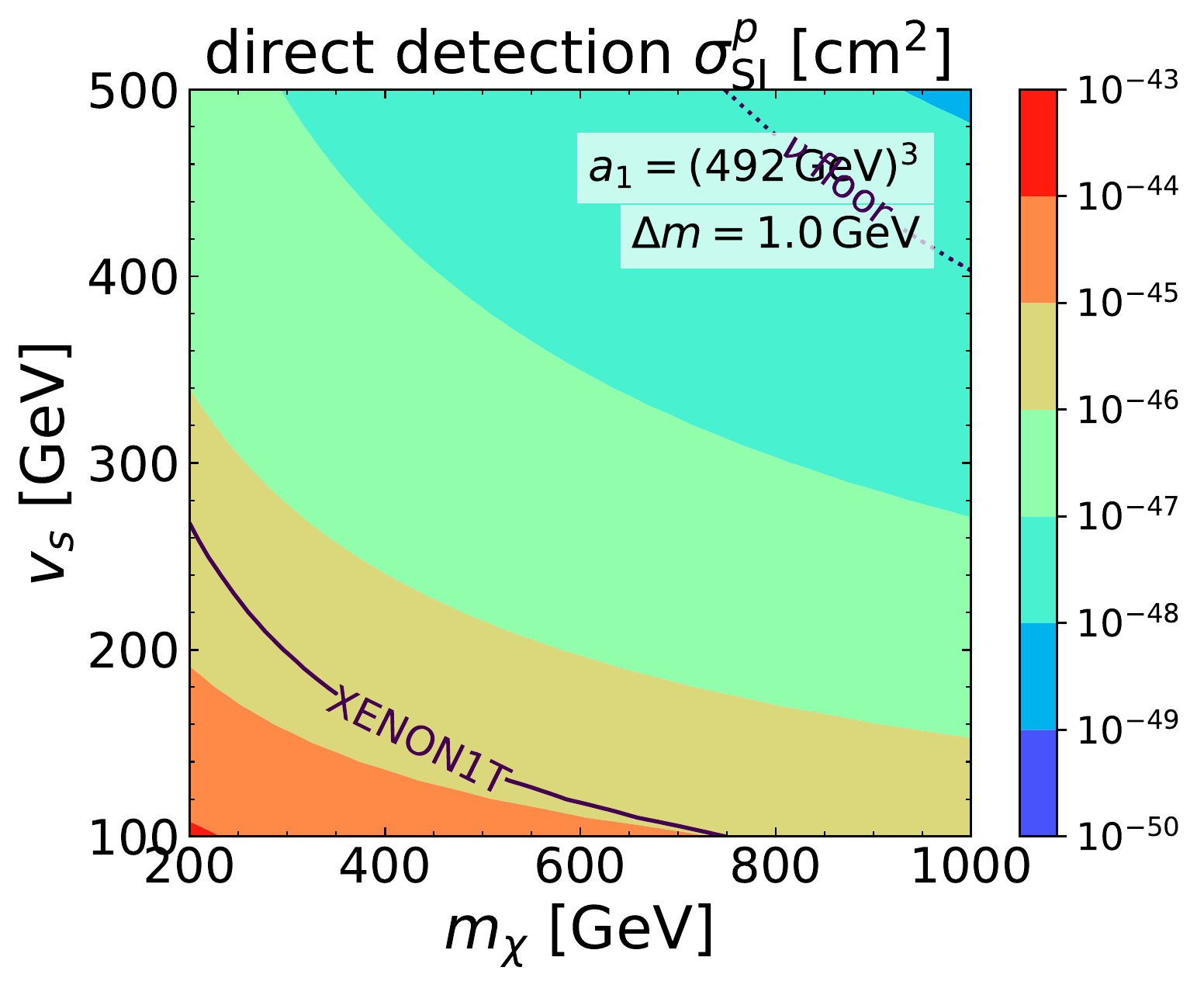}
      \end{minipage}
\\
      \begin{minipage}[t]{0.3\hsize}
        \centering
        \includegraphics[keepaspectratio, scale=0.33]{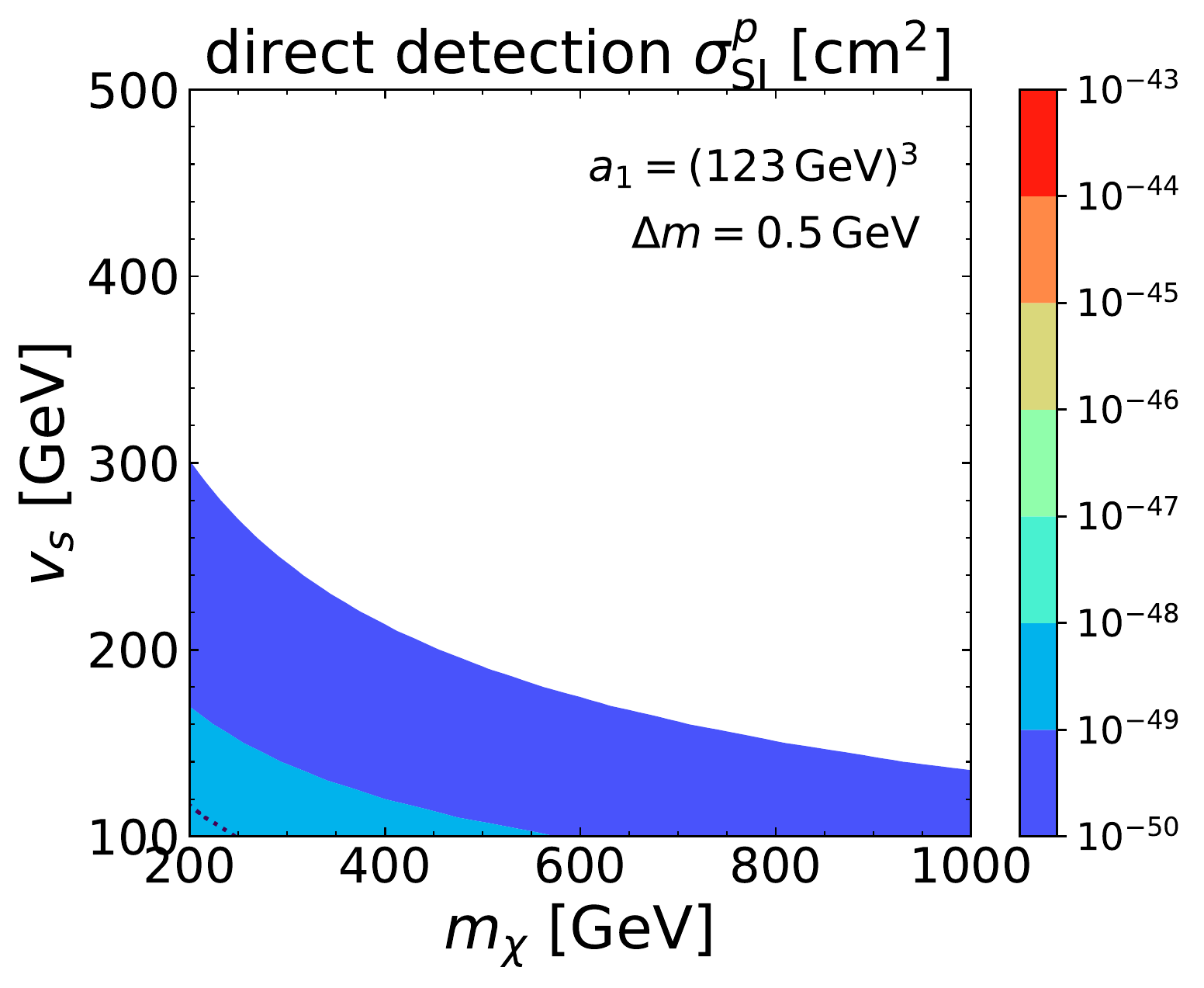}
      \end{minipage} &
      \begin{minipage}[t]{0.3\hsize}
        \centering
        \includegraphics[keepaspectratio, scale=0.33]{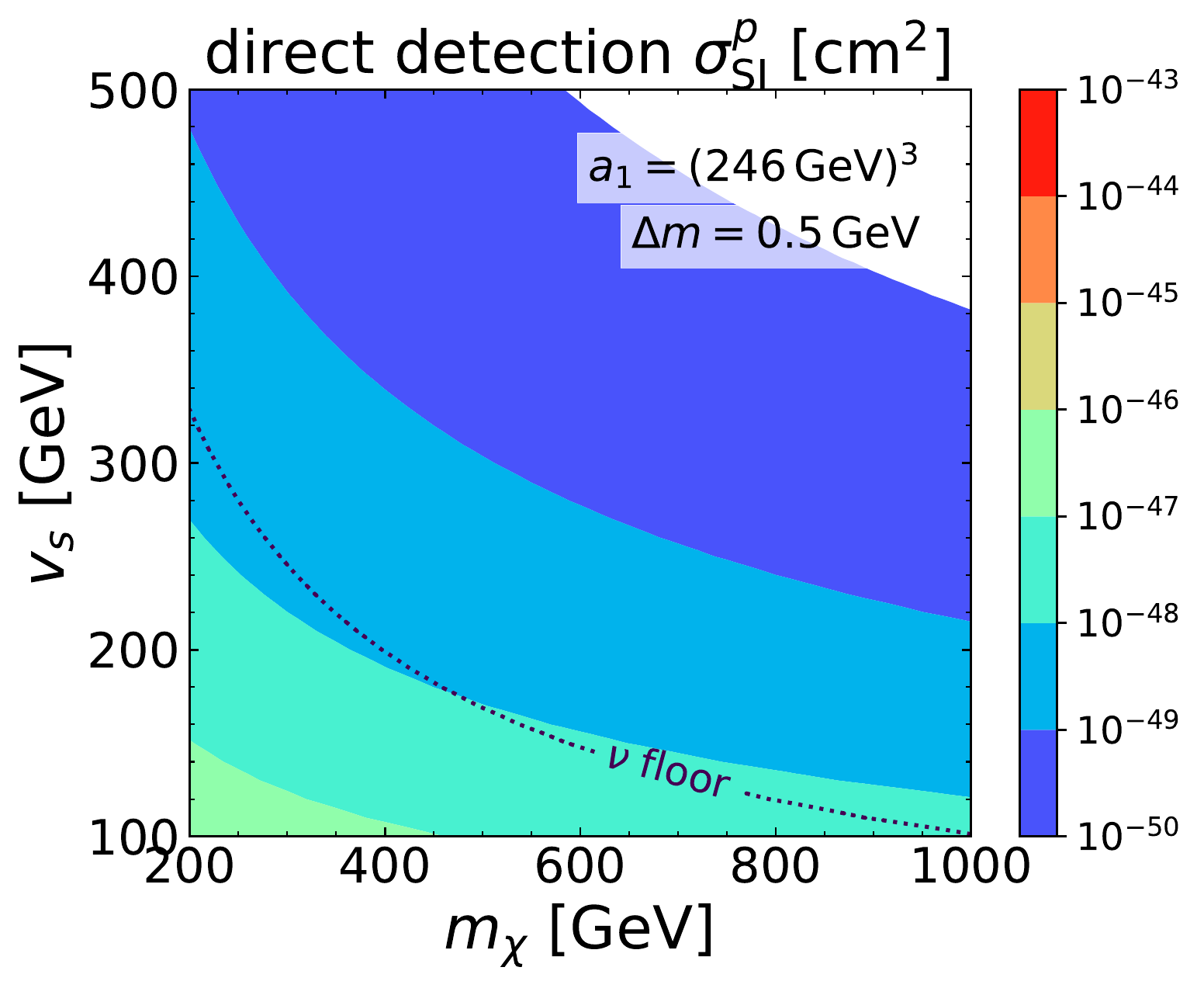}
      \end{minipage} &
      \begin{minipage}[t]{0.3\hsize}
        \includegraphics[keepaspectratio, scale=0.33]{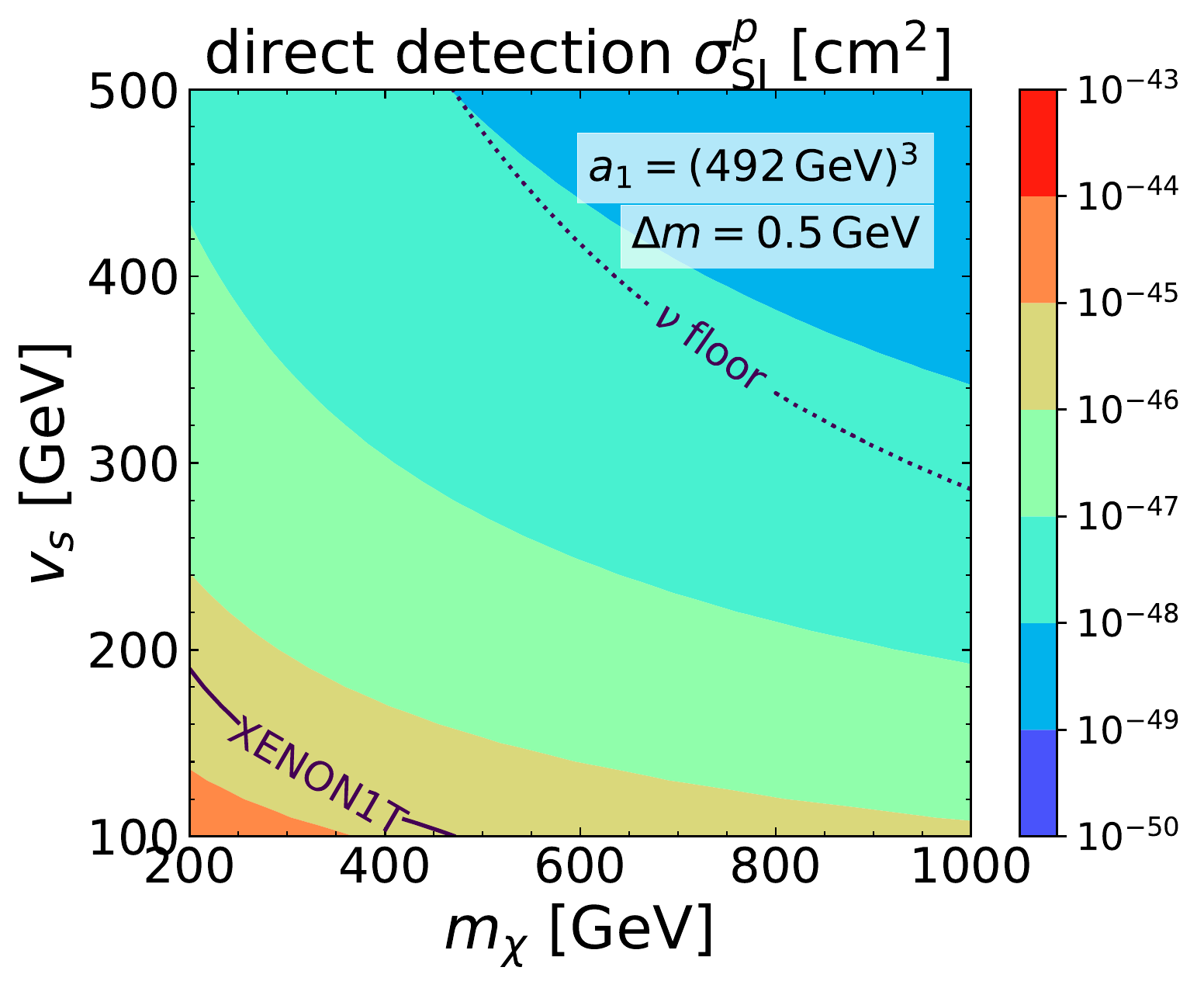}
      \end{minipage}
\\
      \begin{minipage}[t]{0.3\hsize}
        \centering
        \includegraphics[keepaspectratio, scale=0.33]{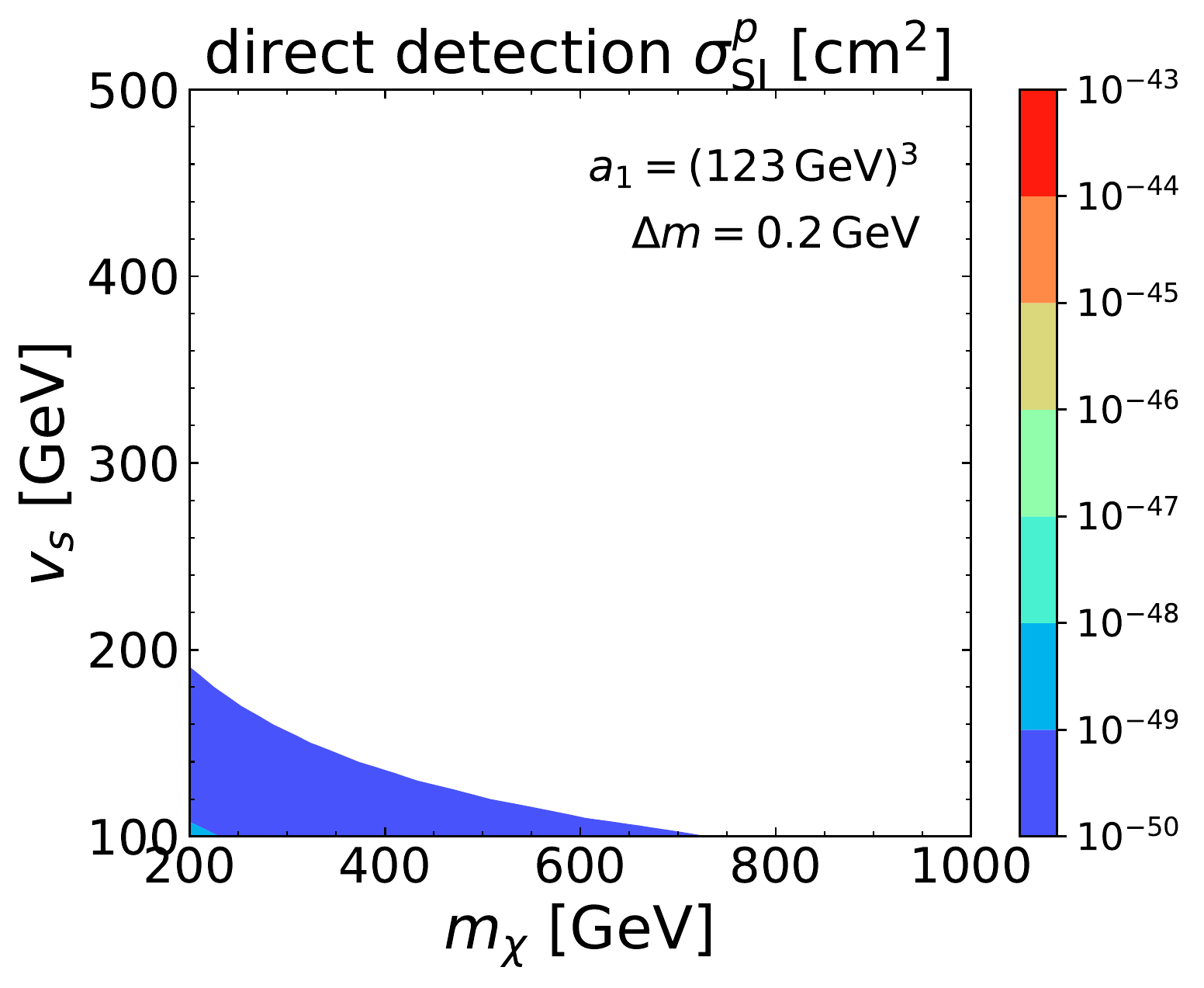}
      \end{minipage} &
      \begin{minipage}[t]{0.3\hsize}
        \centering
        \includegraphics[keepaspectratio, scale=0.33]{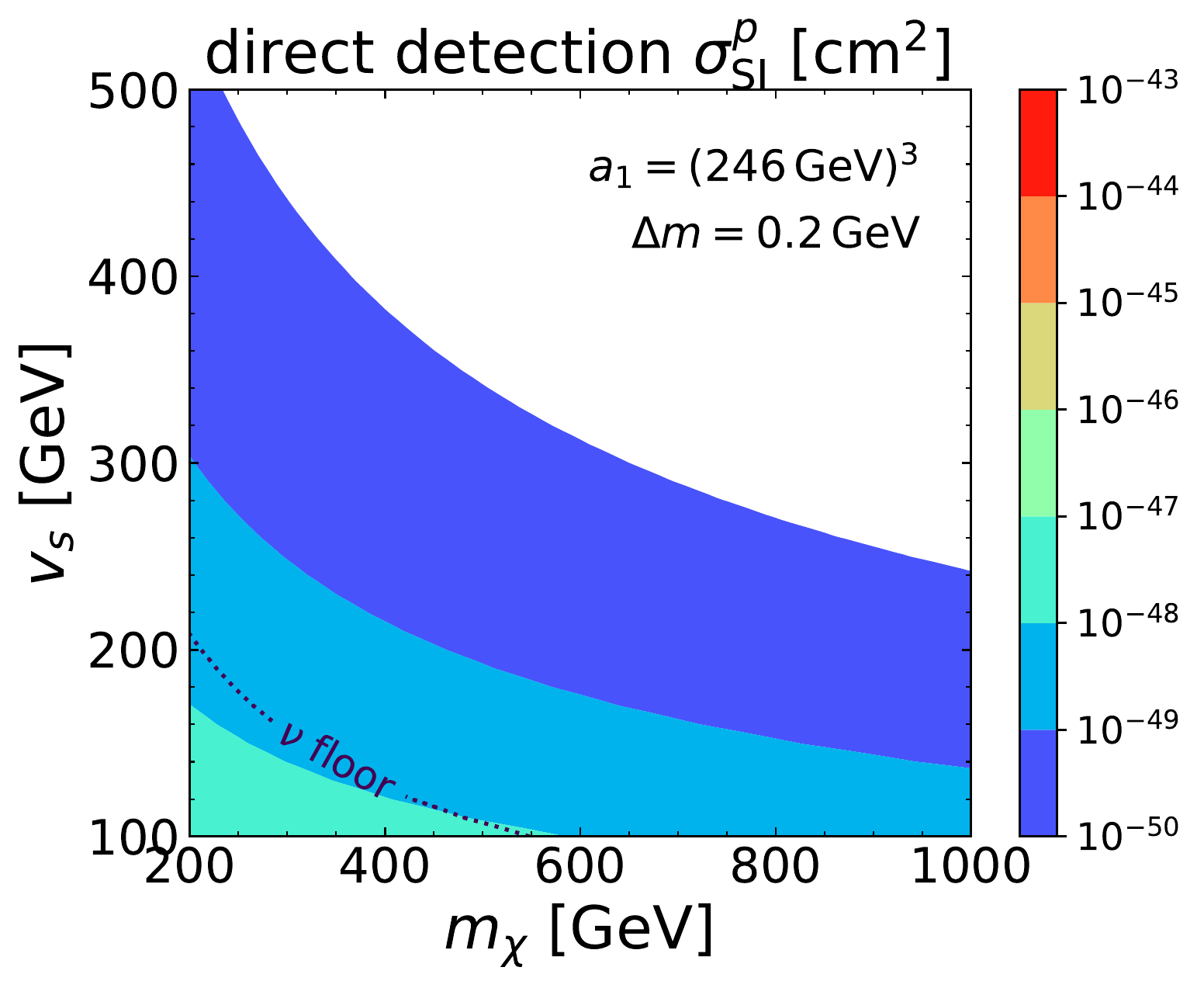}
      \end{minipage} &
      \begin{minipage}[t]{0.3\hsize}
        \includegraphics[keepaspectratio, scale=0.33]{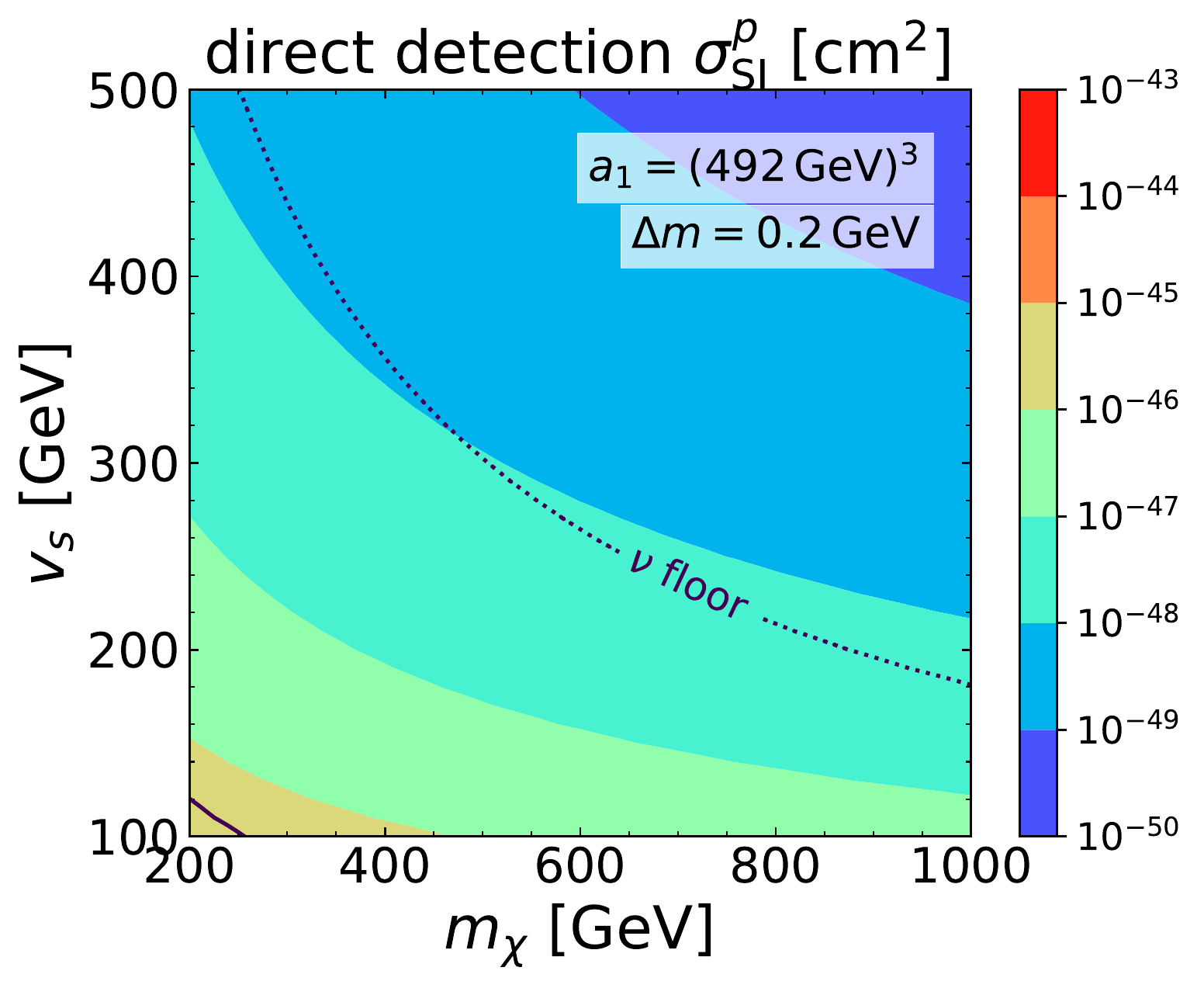}
      \end{minipage}
    \end{tabular}	
 \caption{
 Spin-independent DM--nucleon scattering cross section on the $m_\chi$--$v_s$ plane
 for $\Delta m=1.0,\,0.5,\,0.2$~GeV from the top to bottom panels,
 where $a_1$ is set as $(123~{\rm GeV})^3$, $(246~{\rm GeV})^3$, $(492~{\rm GeV})^3$
 from the left to right panels.
 }
\label{fig:dd_dm}
\end{figure}

The second and third rows of Fig.~\ref{fig:dd} show 
the cross section $\sigma^p_{\mathrm{SI}}$ for 
$v_S=(123,\,246,\,492)~\mathrm{GeV}$ and 
$m_\chi=(200,\,600, \,1000)~\mathrm{GeV}$, respectively. 
The common feature of these graphs is that the cross section increases as $a_1$ increases, 
as already mentioned above.
The larger $m_\chi$ and $v_s$ are, the smaller the cross section becomes.
We note that the theoretical constraints in eqs.~\eqref{th_per} and \eqref{th_stab}
are also satisfied on the entire parameter space presented in Fig.~\ref{fig:dd}.

In Fig.~\ref{fig:dd_dm} 
we show the $\Delta m$-dependence of the cross section $\sigma^p_{\mathrm{SI}}$
on the $m_\chi$--$v_S$ plane. 
We adopt cases in the first row of Fig.~\ref{fig:dd} as examples 
to discuss the $\Delta m$-dependence so that we list them on the first row of Fig.~\ref{fig:dd_dm}.  
Three different values of $\Delta m$ is chosen as 
$\Delta m=1~\mathrm{GeV}$ (top), $0.5~\mathrm{GeV}$ (middle) and $0.2~\mathrm{GeV}$ (bottom),
and $a_1$ is $(123~\mathrm{GeV})^3$ (left), $(246~\mathrm{GeV})^3$ (center) 
and $(492~\mathrm{GeV})^3$ (right). 
While the degree of the degeneracy between the two scalars is increasing,    
the cross section drops sharply, as already seen in Fig.~\ref{fig:fig_DM_mh}.
For $\Delta m=0.2$~GeV, 
the whole parameter space considered in Fig.~\ref{fig:dd_dm} is no longer constrained,
except a tiny corner of the small $m_\chi$ and $v_S$ for $a_1=(492~\mathrm{GeV})^3$.

\begin{figure}
    \begin{tabular}{ccc}
      \begin{minipage}[t]{0.3\hsize}
        \centering
        \includegraphics[keepaspectratio, scale=0.33]{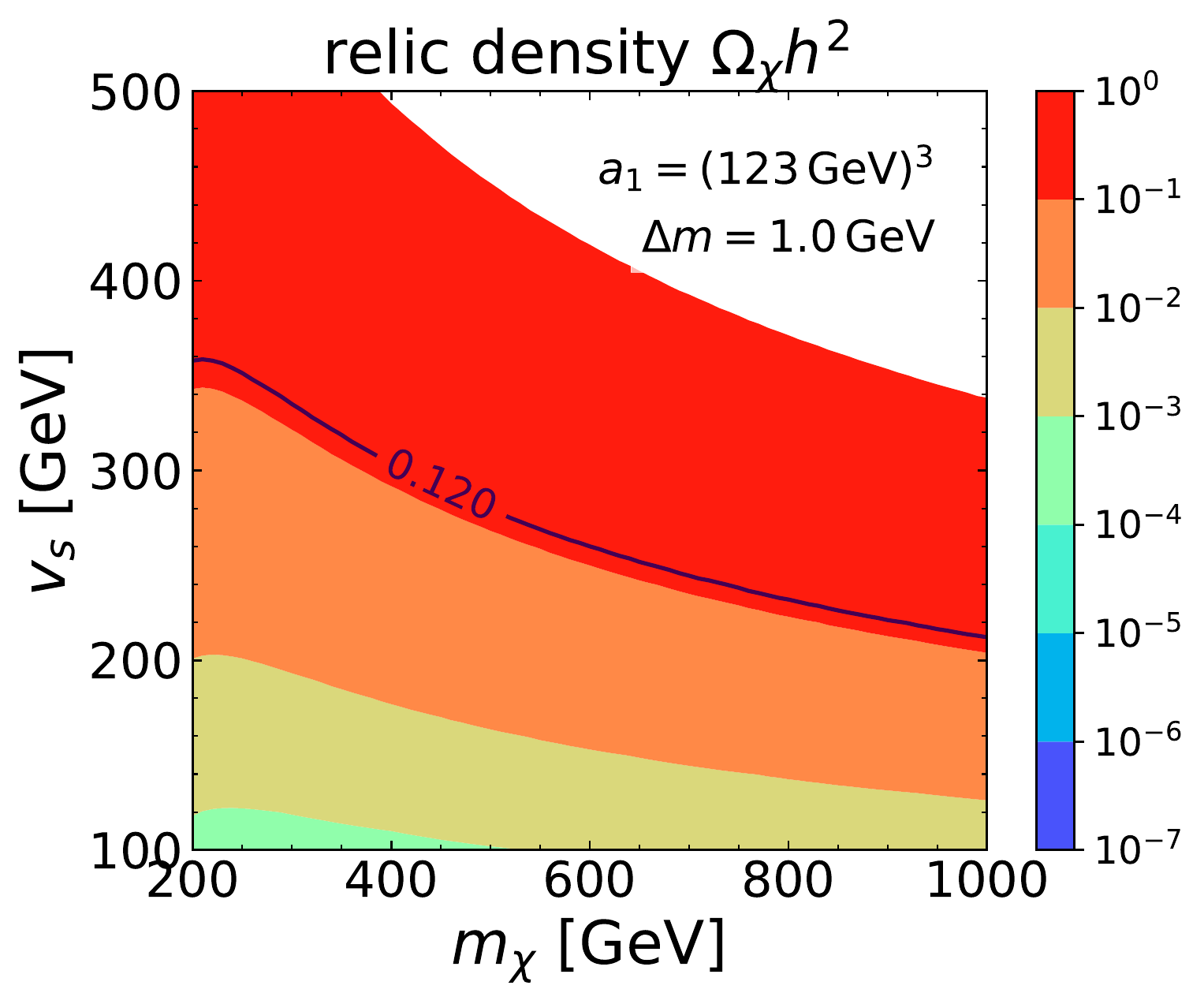}
      \end{minipage} &
      \begin{minipage}[t]{0.3\hsize}
        \centering
        \includegraphics[keepaspectratio, scale=0.33]{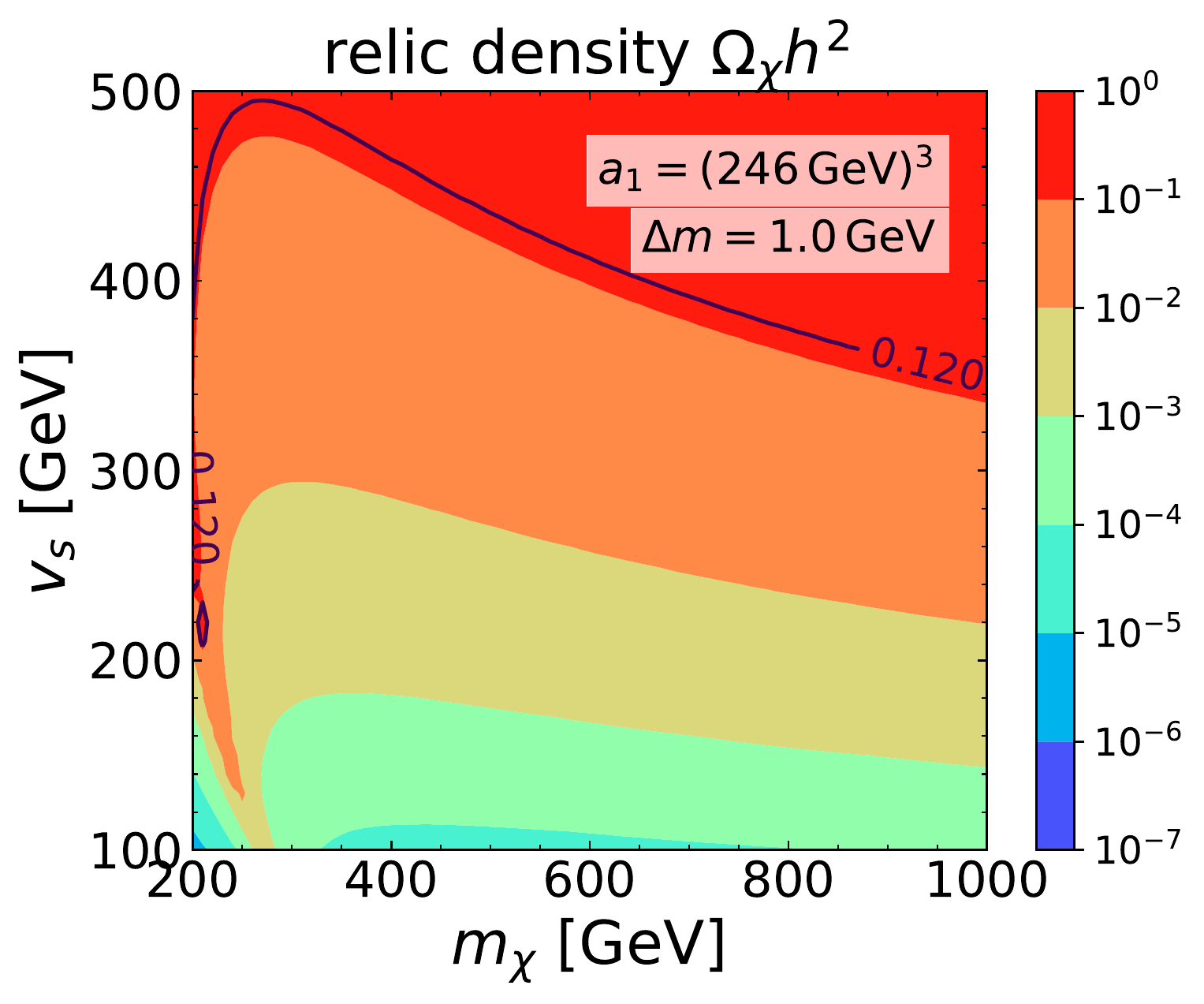}
      \end{minipage} &
      \begin{minipage}[t]{0.3\hsize}
        \includegraphics[keepaspectratio, scale=0.33]{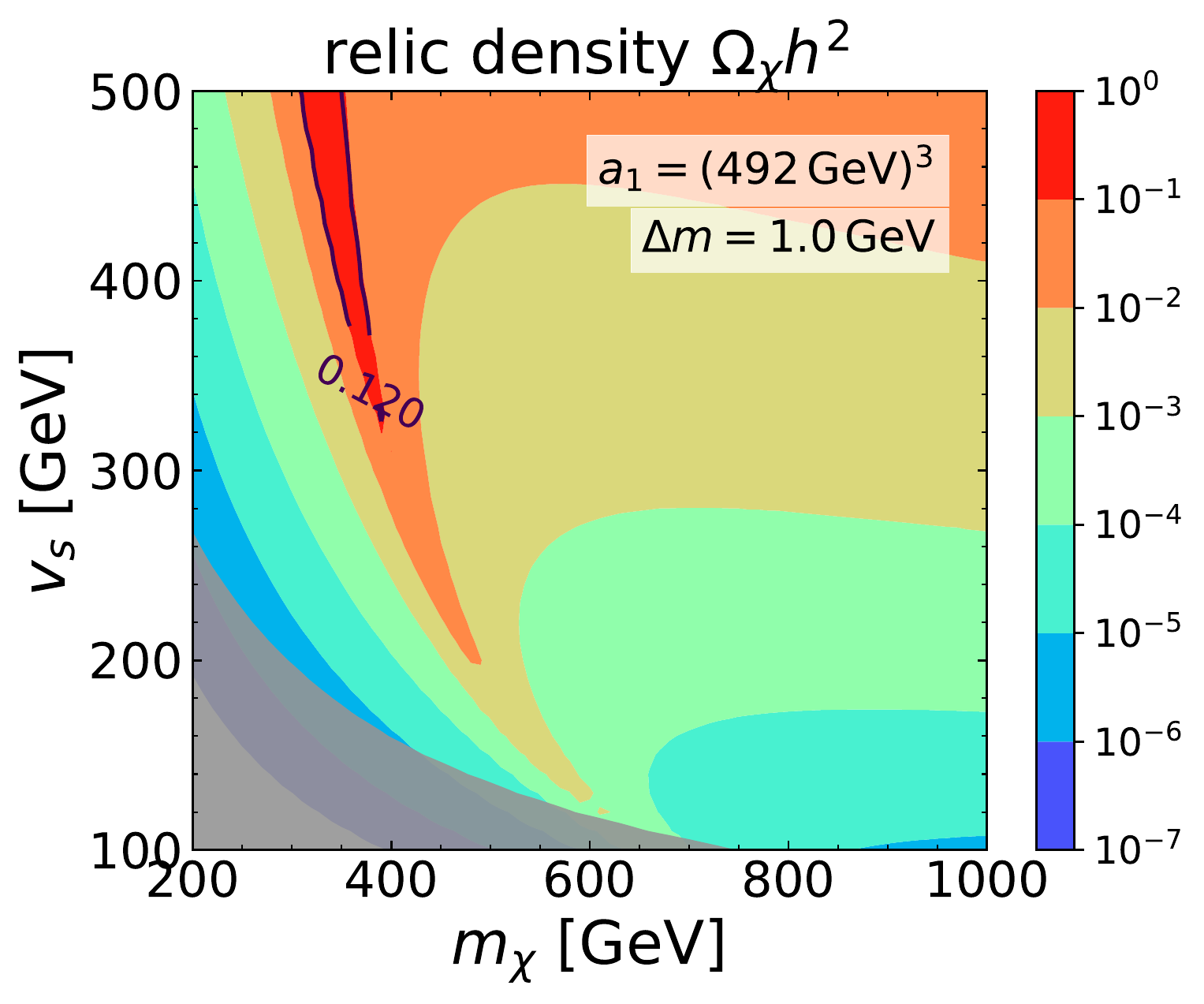}
      \end{minipage}
\\
      \begin{minipage}[t]{0.3\hsize}
        \centering
        \includegraphics[keepaspectratio, scale=0.33]{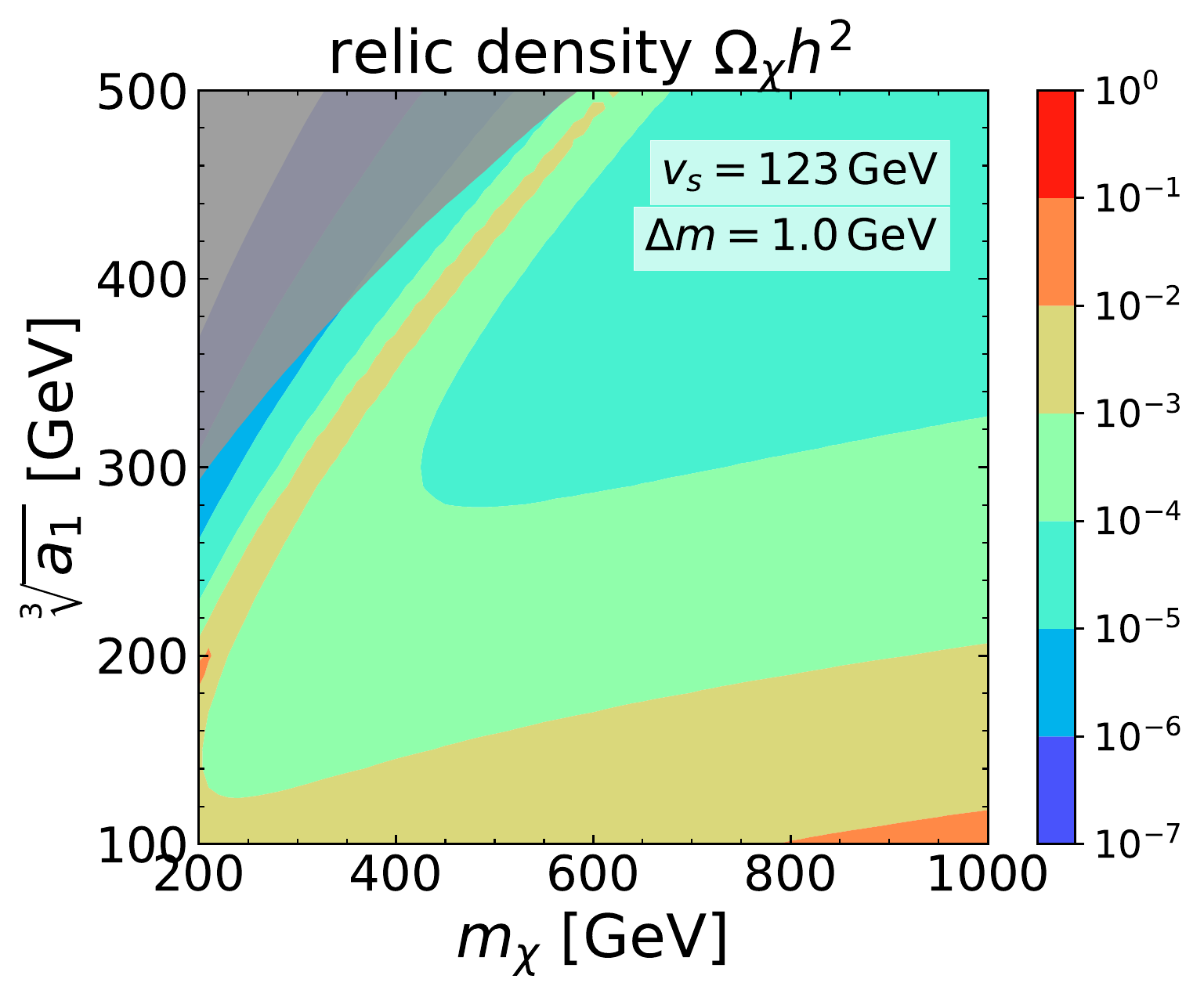}
      \end{minipage} &
      \begin{minipage}[t]{0.3\hsize}
        \centering
        \includegraphics[keepaspectratio, scale=0.33]{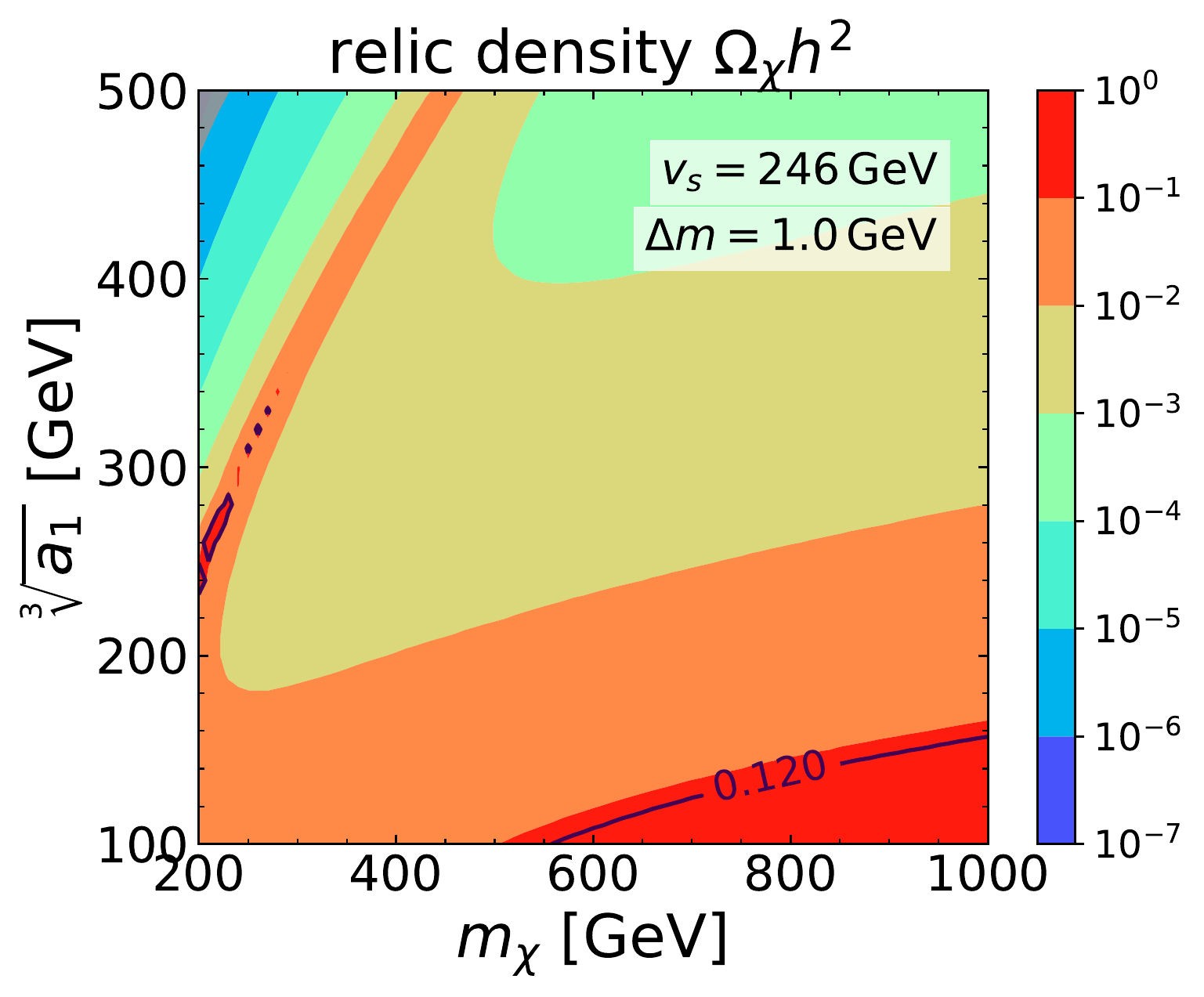}
      \end{minipage} &
      \begin{minipage}[t]{0.3\hsize}
        \includegraphics[keepaspectratio, scale=0.33]{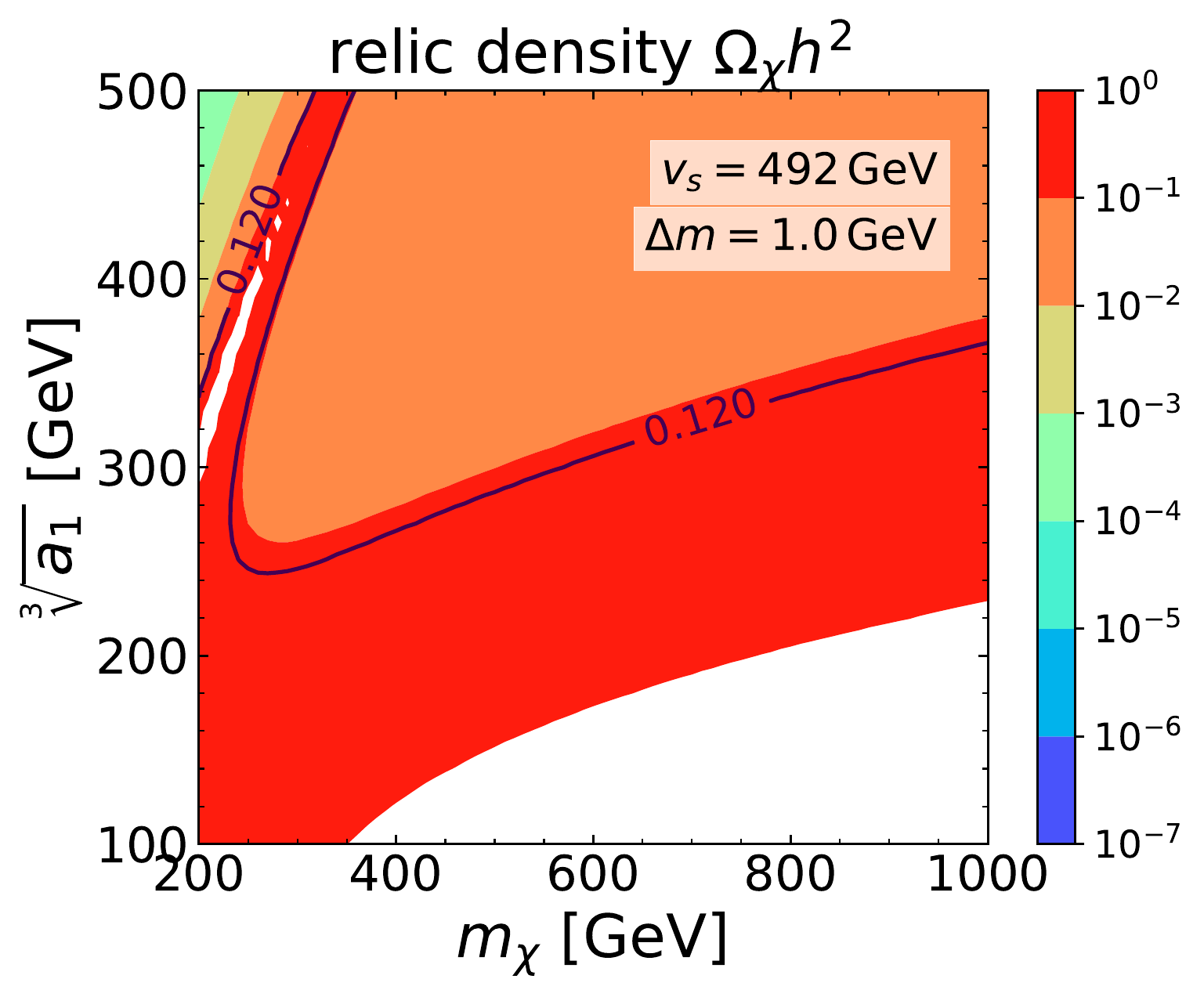}
      \end{minipage}
\\
      \begin{minipage}[t]{0.3\hsize}
        \centering
        \includegraphics[keepaspectratio, scale=0.33]{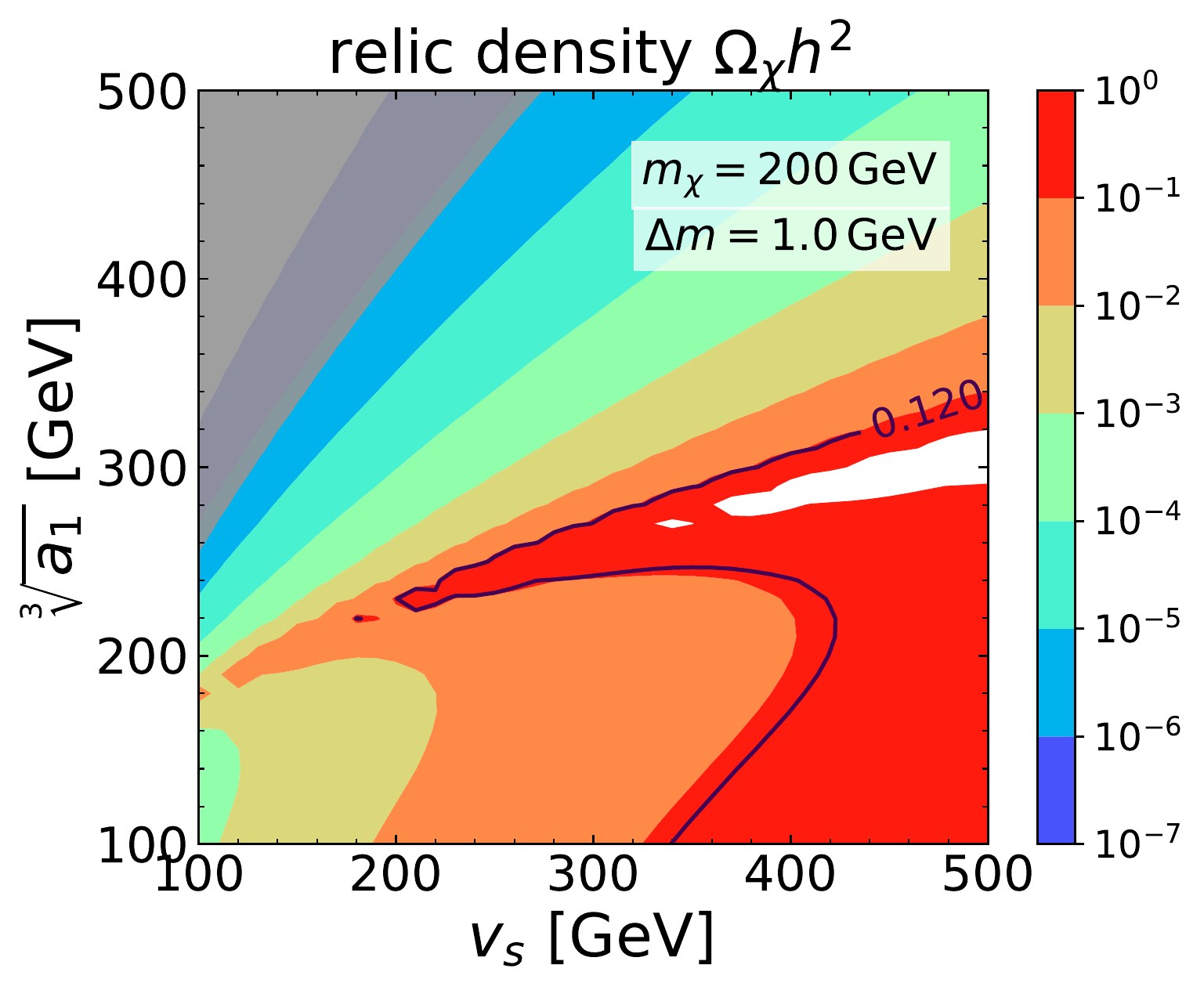}
      \end{minipage} &
      \begin{minipage}[t]{0.3\hsize}
        \centering
        \includegraphics[keepaspectratio, scale=0.33]{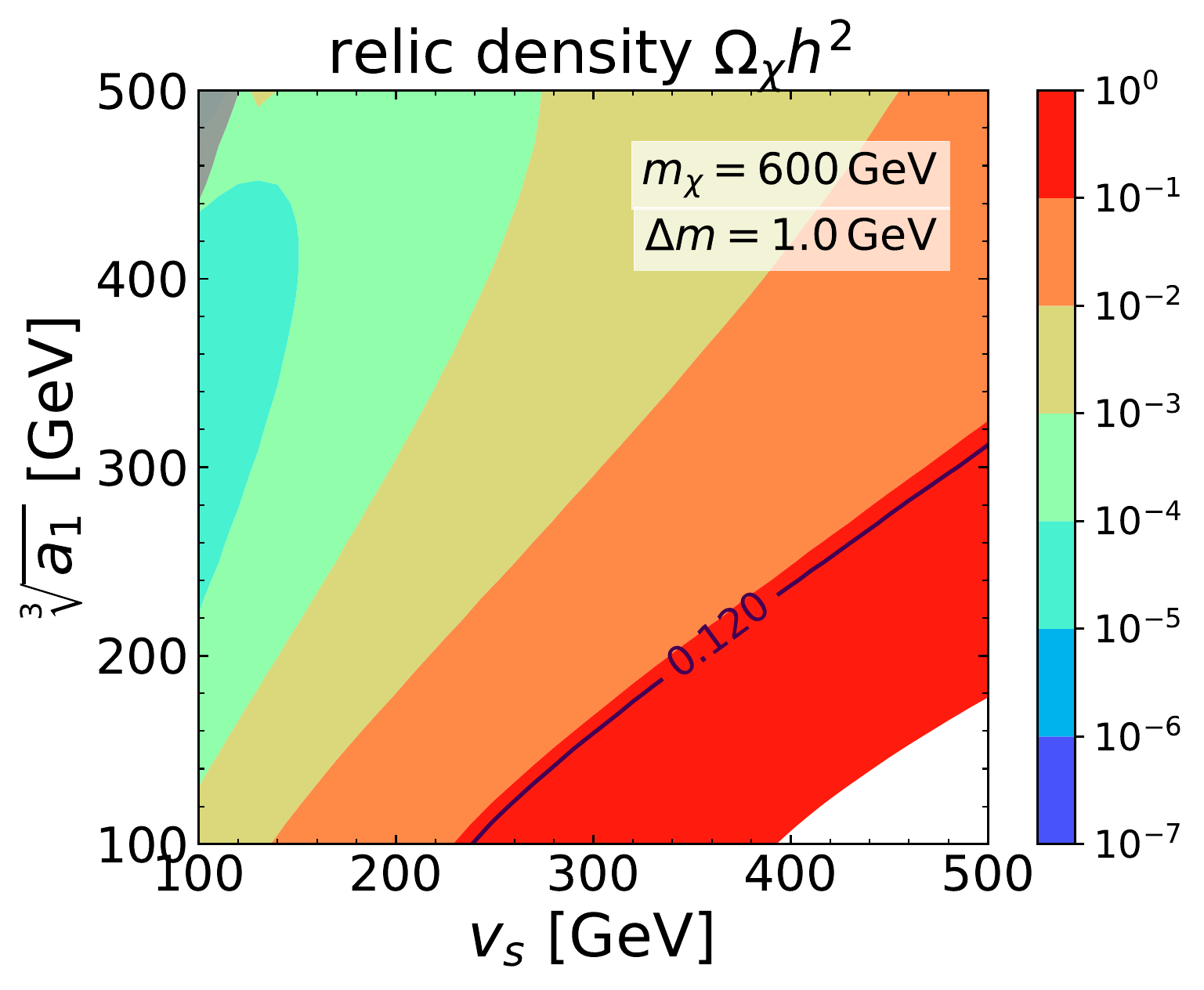}
      \end{minipage} &
      \begin{minipage}[t]{0.3\hsize}
        \includegraphics[keepaspectratio, scale=0.33]{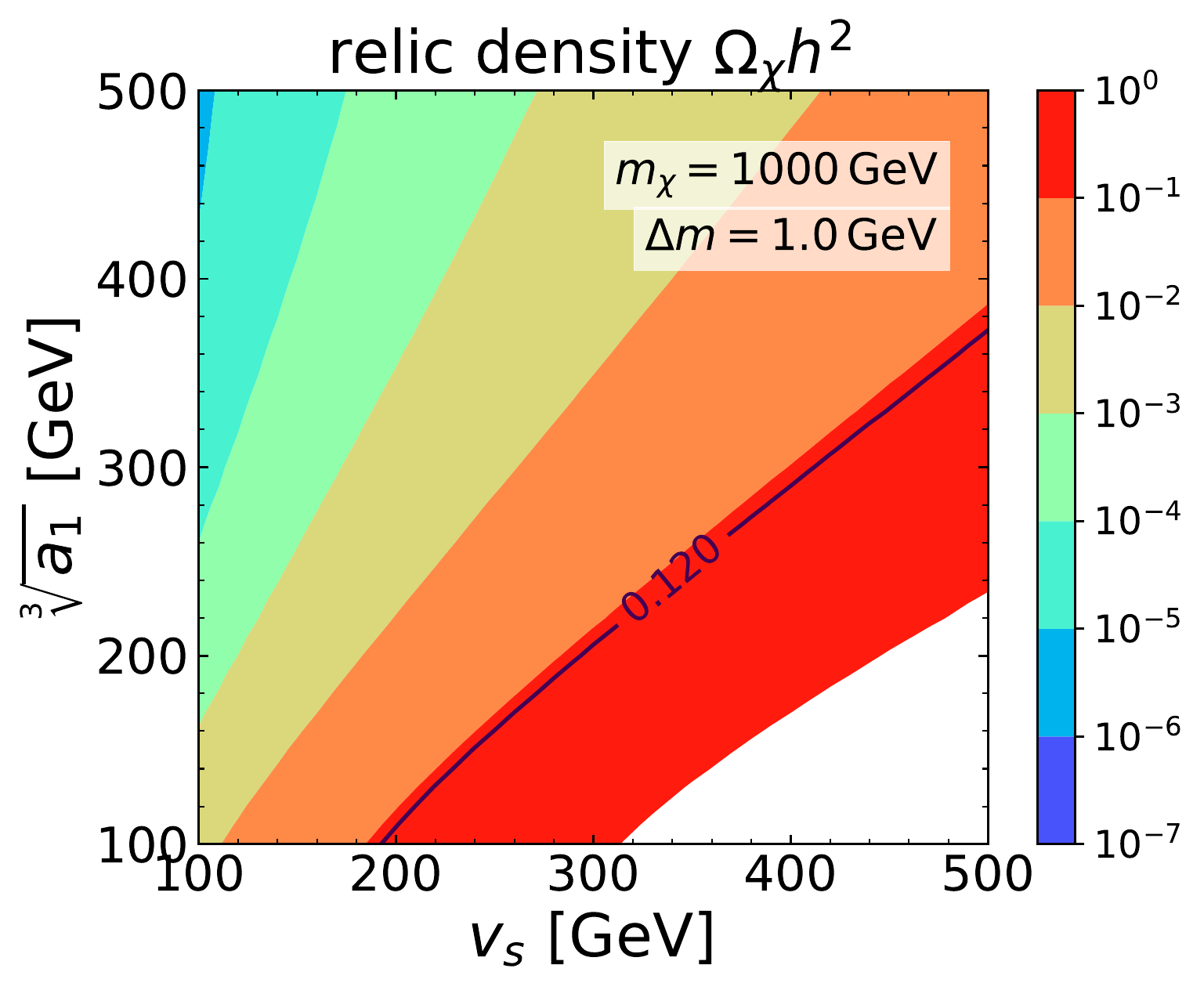}
      \end{minipage}
    \end{tabular}
 \caption{
 DM relic density on the $m_\chi$--$v_s$ (top),
 $m_\chi$--$\sqrt[3]{a_1}$ (middle), and $v_s$--$\sqrt[3]{a_1}$ (bottom) planes.
 $\Delta m=m_{h_2}-m_{h_1}$ is fixed at 1~GeV for all the panels, 
 while the other parameter, $a_1$ (top), $v_s$ (middle), and $m_\chi$ (bottom), 
 is taken as the low to high scale from the left to right panels.
 Gray shaded regions denote the region excluded by the XENON1T direct detection experiment.
}
\label{fig:relic}
\end{figure}

Up to now, we have studied the parameter space of the pseudoscalar DM model
with nearly degenerate scalars 
and found that sizable parameter regions are allowed from the direct detection experiments. 
In Fig.~\ref{fig:relic} we here show constraints on parameters $(m_\chi,\,v_s,\,a_1)$ 
from the relic density of the DM $\Omega_\chi h^2$.
In the figure, contour plots of $\Omega_\chi h^2$ on 
$m_\chi$--$v_S$, $m_\chi$--$\sqrt[3]{a_1}$ and $v_s$--$\sqrt[3]{a_1}$ planes 
are shown in the first, second and third rows, respectively. 
The parameter choices are same as in Fig.~\ref{fig:dd}.
We note that, although we set $\Delta m=1$~GeV as a benchmark, 
the results are not altered for $\Delta m=0.5~\mathrm{GeV}$ or $0.2~\mathrm{GeV}$,
different from the DM--nucleon scattering cross section as shown in Fig.~\ref{fig:dd_dm}. 
The black curve denotes $\Omega_\chi h^2=0.120$ from the Planck measurement~\cite{Aghanim:2018eyx},
and the region of $\Omega_\chi h^2>0.120$ is disfavored
since the DM is overproduced in this model. 

From the first row in Fig.~\ref{fig:relic}, 
we see that the relic density $\Omega_\chi h^2$ could be large as $v_S$ increases,
while small as $a_1$ increases.
This can be explained as follows.
The dominant annihilation processes are $\chi \chi \to h_i h_j~(i,j=1,2)$
since the processes for other final states ($qq,\,\ell\ell,\,VV$) are 
strongly suppressed by the same mechanism with the case of the DM--quark scattering $\chi q \to \chi q$, 
where $q,\ell$ and $V$ stand for quarks, leptons and vector bosons, respectively. 
The scalar trilinear interactions for $h_1$-$\chi$-$\chi$ and $h_2$-$\chi$-$\chi$
in eq.~\eqref{trilinear} are proportional to a ratio $a_1/v_S$.
Therefore, larger $v_S$ suppresses the annihilation rate of the DM,  
while for larger $a_1$ the scalar trilinear interactions become stronger
and the annihilation of the DM pair is enhanced. 
The similar behaviors can be observed in the second and third rows,
i.e. on the different parameter planes. 

One can see a strip in some particular parameter regions,
where the relic density suddenly becomes large,
e.g. for $m_\chi=300-400$~GeV on the top--right panel in Fig.~\ref{fig:relic}.
This can be explained by the suppression of the annihilation cross section
due to a non-trivial cancellation among the $\chi\chi\to h_ih_j$ amplitudes;
the four-point, $s$-channel $h_i$-mediated, and $t$-channel $\chi$-mediated amplitudes.

In Fig.~\ref{fig:relic}, we overlay the exclusion regions
from the XENON1T direct detection experiment, shown in Fig.~\ref{fig:dd}.
The exclusion regions from the direct detections tend to correspond to the regions
where the relic density is very small, and vice versa. 
The wide parameter regions of $(m_\chi,v_s,\sqrt[3]{a_1})$ are still unconstrained 
from both the direct DM detection experiments and the measurement of the DM relic density.
If a DM signal at direct detection experiments is observed in near future, 
the parameter space in this model can be narrowed down. 

\begin{figure}
 \center
 \includegraphics[width=0.33\textwidth]{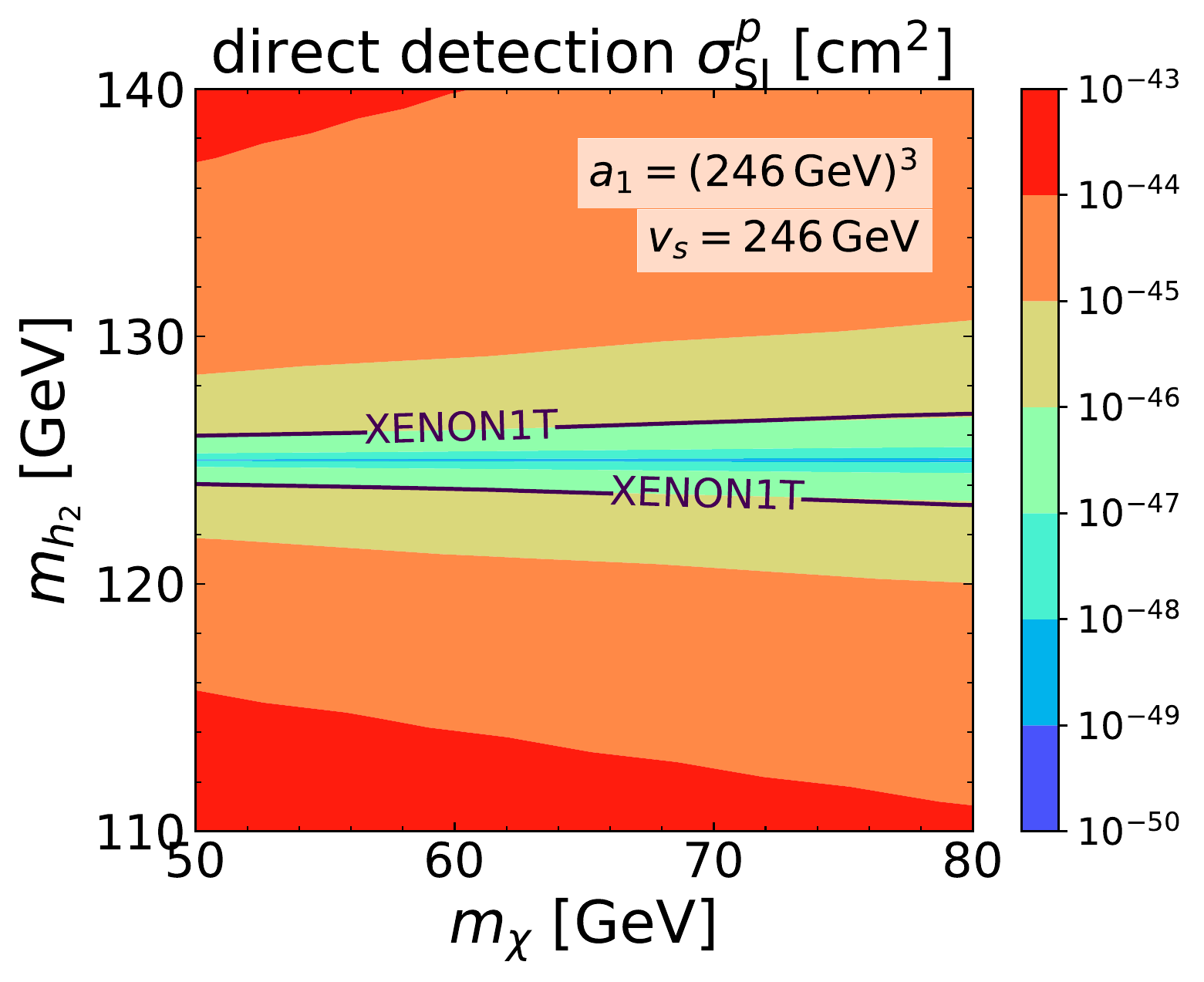} \qquad
 \includegraphics[width=0.33\textwidth]{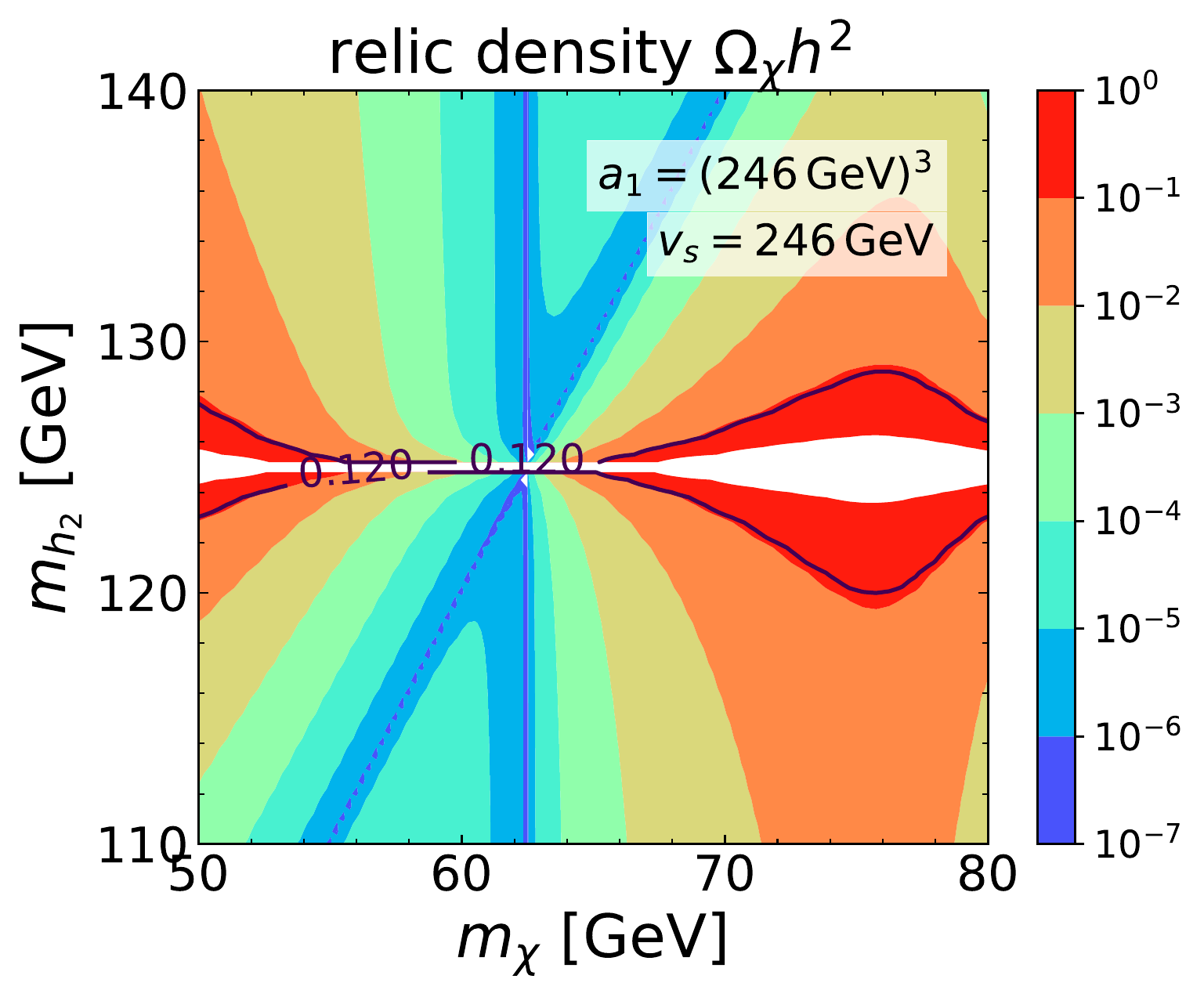}
 \caption{
  Spin-independent DM--nucleon scattering cross section (left) and DM relic density (right)
  on the $m_\chi$--$m_{h_2}$ plane for the low $m_\chi$ region.
 }
 \label{fig:lowDMmass}
\end{figure}

Before closing this section, we briefly mention a case for the low DM mass, especially for
around $m_\chi\sim m_{h_1}/2$.
In Fig.~\ref{fig:lowDMmass},  
we show contour plots of the spin-independent cross section $\sigma^p_{\mathrm{SI}}$ (left)
and the relic density of the DM $\Omega_\chi h^2$ (right) 
on $m_\chi$--$m_{h_2}$ plane
for $v_S=\sqrt[3]{a_1}=246~\mathrm{GeV}$. 
As for the DM--nucleon scattering cross section, the suppression around $m_{h_2}\sim m_{h_1}$ can
also be seen in this low $m_\chi$ region, while the $m_\chi$ dependence is very mild.
On the other hand, the relic density strongly depends on $m_\chi$ as well as $m_{h_2}$.
As seen, in this low $m_\chi$ case, the degenerate scalar scenario, $m_{h_2}\sim m_{h_1}$, is excluded by the limit from
the relic density. 
This is because, different from the case shown in Fig.~\ref{fig:relic}, 
the DM annihilation to the fermions and the gauge bosons are dominant,
which are strongly suppressed in this scenario.
Apart from the case of $m_{h_2}\sim m_{h_1}$, 
especially for $m_\chi\sim m_{h_1}/2$ or $m_\chi\sim m_{h_2}/2$,
the relic density becomes very low 
since the DM annihilation is enhanced by the scalar resonance, which can be commonly observed   
in Higgs-portal DM models~\cite{Arcadi:2019lka}.

\section{Test for a degenerate-scalar scenario at the ILC}\label{coll}

We have so far discussed that the stringent constraints on DM models from direct detection experiments
could be avoided in the pseudoscalar DM model with nearly degenerate scalars
without contradicting the observed relic density of the DM
as well as theoretical constraints on the scalar potential.
In this section we discuss possibilities to verify such a degenerate-scalar scenario in collider experiments.

A typical DM signature at colliders is missing energy from DM production. 
In the pseudoscalar DM model, the DM $\chi$ (the CP-odd scalar) only couples
to the CP-even scalars $h_1$ and $h_2$ (the SM-like 125~GeV Higgs boson and the other scalar).
Therefore, unless $m_{\chi}>m_{h_{1,2}}/2$, 
the $h_{1,2}$ productions followed by their decays into a pair of $\chi$
can lead to such missing-energy signature.
Especially, the degenerate-scalar case can be observed as invisible Higgs decay.
We note that in such a case it is very difficult to distinguish the model 
from a simpler Higgs portal DM model, 
where DM only couples to the single SM-like Higgs boson.
In the following, however, we will show that it is possible to examine the degenerate-scalar scenario using the process other than the invisible decay. 

In order to test a degenerate-scalar scenario, instead of the missing-energy signature,
here we focus on degenerate scalar productions at the ILC~\cite{Behnke:2013xla},
and investigate how well we can distinguish the degenerate states
from the single state of $m_h=125$~GeV in the SM.
We note that, at the Large Hadron Collider (LHC),
by using the high resolution of the diphoton channel of the Higgs boson decays, 
the mass difference between the two degenerate states $\Delta m\gtrsim 3$~GeV
is disfavored at the 2$\sigma$ level from the LHC Run-I data~\cite{Khachatryan:2014ira}.
Although no such specific analyses can be found in refs.~\cite{Aaboud:2018wps,Sirunyan:2020xwk}
with the LHC Run-II data, we expect a better resolution. 
The phenomenological studies on mass-degenerate Higgs bosons are also found
in, e.g., refs.~\cite{Gunion:2012he,Ferreira:2012nv,Robens:2015gla,Bian:2017gxg}.

The main target of the ILC at $\sqrt{s}=250$~GeV (ILC250) is 
Higgs boson production associated with a $Z$ boson~\cite{Fujii:2017vwa}.
The mass of the Higgs boson is precisely determined
by the recoil mass technique in the process~\cite{Yan:2016xyx}
\begin{align}
	e^+e^-\to h_{1,2}Z\to h_{1,2}\ell^+\ell^-,
\label{process}
\end{align}
where the Higgs mass is reconstructed from $\ell=e$ or $\mu$ as 
\begin{align}
	m^2_{\rm recoil}=(\sqrt{s}-E_{\ell\ell})^2-|\vec{p}_{\ell\ell}|^2.
\end{align}
The recoil mass is independent of how the degenerate scalars decay,
and hence this analysis is independent of the mass hierarchy
between the DM and the degenerate scalars. 
We note that, different from processes where the DM involves
such as DM--nucleon scattering and DM annihilation, 
there is no cancellation between the two amplitudes mediated by $h_1$ and $h_2$ in the SM-like processes.
\footnote{
Even when the $h_{1,2}\to\chi\chi$ decays are allowed in the process~\eqref{process}, unless the mass difference is smaller than the widths of $h_{1,2}$, no cancellation happens.
}
In the following, we parameterize the mass difference of the two scalars as
\begin{align}
	m_{h_1,h_2}= \Big(125\pm \frac{\Delta m}{2}\Big)~{\rm GeV}. 
\end{align}

\begin{figure}
 \center
 \includegraphics[clip, width=0.5\textwidth]{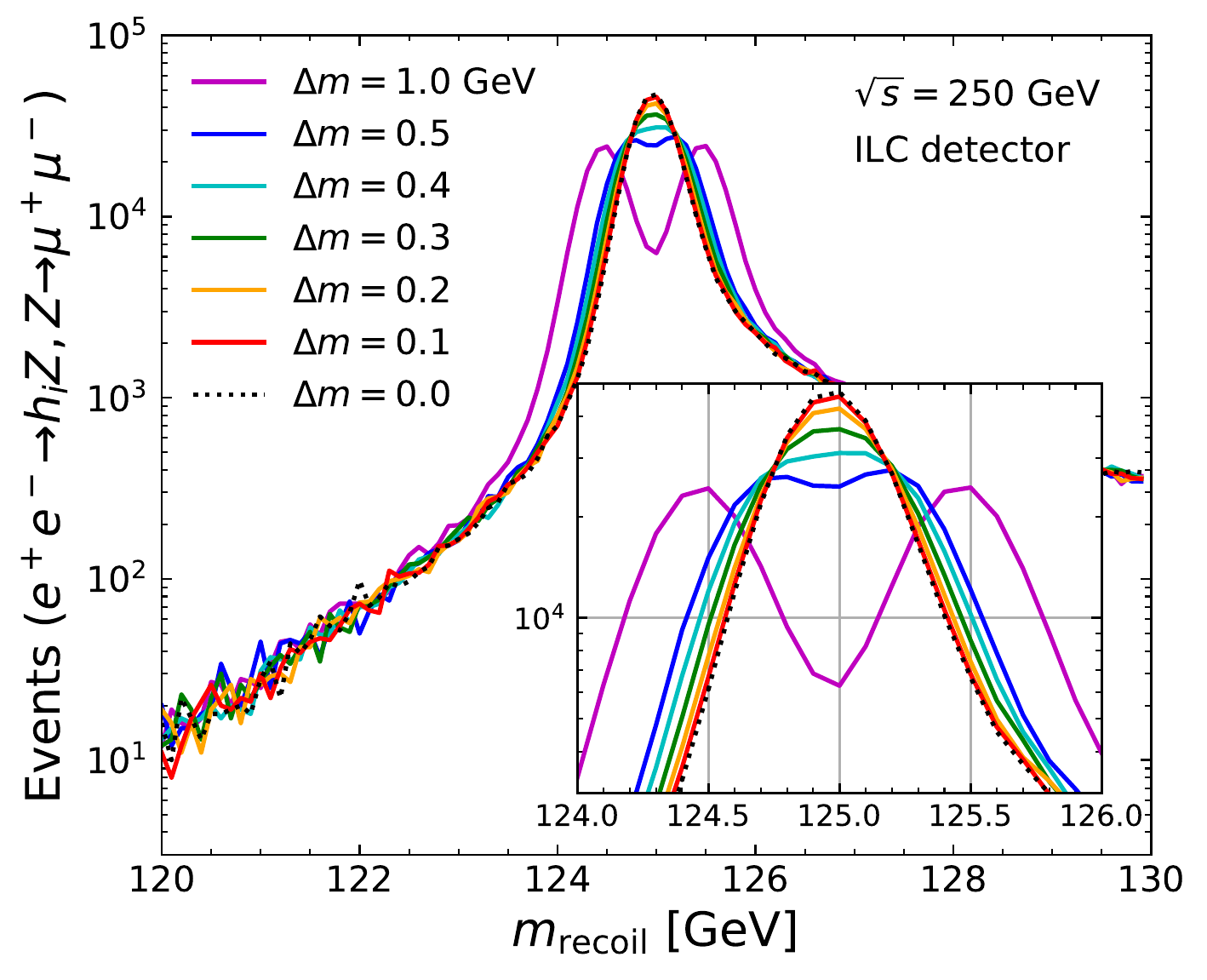}
 \caption{ 
  Recoil mass distributions for various $\Delta m$
  for $e^+e^-\to h_{1,2}Z,\, Z\to\mu^+\mu^-$ at $\sqrt{s}=250$~GeV,
  where the generic ILC detector is assumed.
 }
 \label{fig:mrec}
\end{figure}

Figure~\ref{fig:mrec} shows recoil mass distributions with a muon-pair final state
in the pseudoscalar DM model with nearly degenerate scalars for various $\Delta m$
denoted by different colors.
For simulation, using the same CxSM UFO model as in the DM computations, 
we employed {\sc MadGraph5\_aMC@NLO v2.7.3}~\cite{Alwall:2014hca}
with {\sc Pythia 8.2}~\cite{Sjostrand:2014zea} for event generation
and {\sc Delphes v3.4.2}~\cite{deFavereau:2013fsa} to take into account detector effects.
We used the {\sc ILCDelphes} card: a {\sc Delphes} model
describing a parametrized generic ILC detector~\cite{Fujii:2020pxe,ILCDelphes}
based on two types of detectors proposed for the ILC~\cite{Behnke:2013lya},
SiD (Silicon Detector)~\cite{Aihara:2009ad} and ILD (International Large Detector)~\cite{Abe:2010aa}.
The total cross section for the signal,
i.e. the sum of the two resonances, is independent of $\Delta m$,
while the relative strengths for the two resonances depend on the mixing angle $\alpha$.
Here, as before, we assume $\alpha=\pi/4$, i.e. the two resonances have the same signal strengths.

In Fig.~\ref{fig:mrec}, for $\Delta m=1.0$~GeV, we can clearly observe two peaks.    
For $\Delta m<0.5$~GeV, on the other hand, we no longer observe two peaks,
instead see a broader peak than for the single resonance case
denoted by a dotted curve ($\Delta m=0$).
We note that, as expected, the results strongly depend on the resolution of muon momenta 
determined by detectors.%
\footnote{In the early stage of this work, we performed two detector simulations, SiD and ILD, 
implemented as a {\sc Delphes} detector card in the previous ILC study~\cite{Potter:2016pgp}.
The resolution of the recoil mass spectrum, i.e. the muon resolution, for SiD (ILD) 
is slightly worse (better) than that for the generic ILC detector.}  

In order to assess the minimal luminosity to exclude or discover such a degenerate-scalar scenario, 
we perform tests of significance with the $\chi^2$ function defined as
\begin{align}
	\chi^2=\sum_{i=1}^N\frac{(n^i_{}-n^i_{\rm SM})^2}{n^i_{\rm SM}},
\label{chi2}	
\end{align}
where $n^i_{}$ is the number of events in the $i$-th bin expected in our pseudoscalar DM model, 
and $n^i_{\rm SM}$ is the corresponding prediction in the SM.
As the background in the recoil mass distribution is estimated well
in ref.~\cite{Yan:2016xyx}, we simply take the SM Higgs signal as our zero hypothesis. 
We fix the total number of events at a certain integrated luminosity by the total cross section 10~fb 
($=\sigma(e^+e^-\to hZ)\times B(Z\to\mu^+\mu^-)$) for $m_h=125$~GeV at $\sqrt{s}=250$~GeV
with the planned beam polarization $P(e^-,e^+)=(-0.8,0.3)$~\cite{Fujii:2017vwa}. 
We require at least ten events in each bin to count as the degree of freedom
for the $\chi^2$ calculation in eq.~\eqref{chi2}.

\begin{figure}
 \center
 \includegraphics[width=0.325\textwidth]{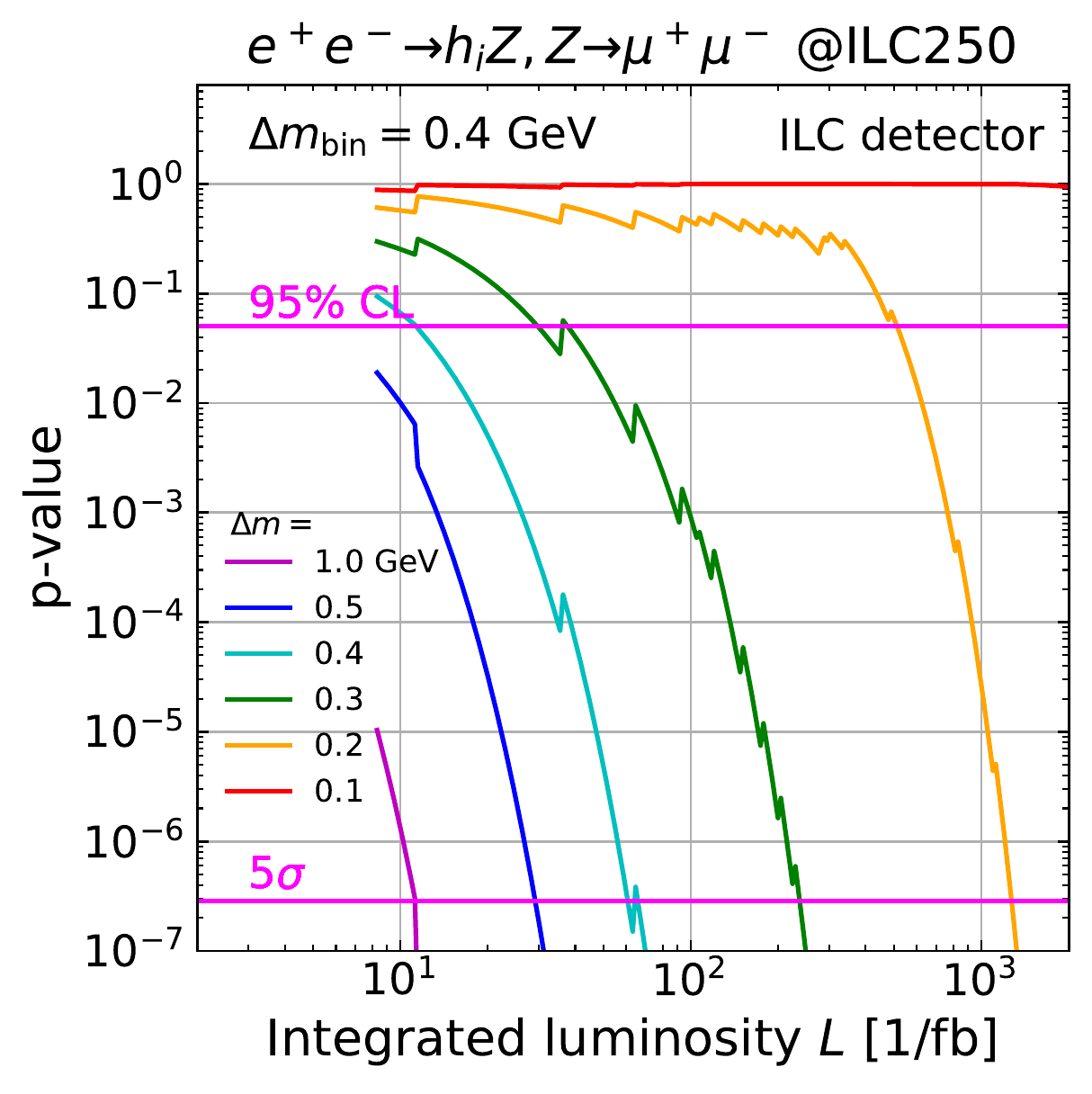}
 \includegraphics[width=0.325\textwidth]{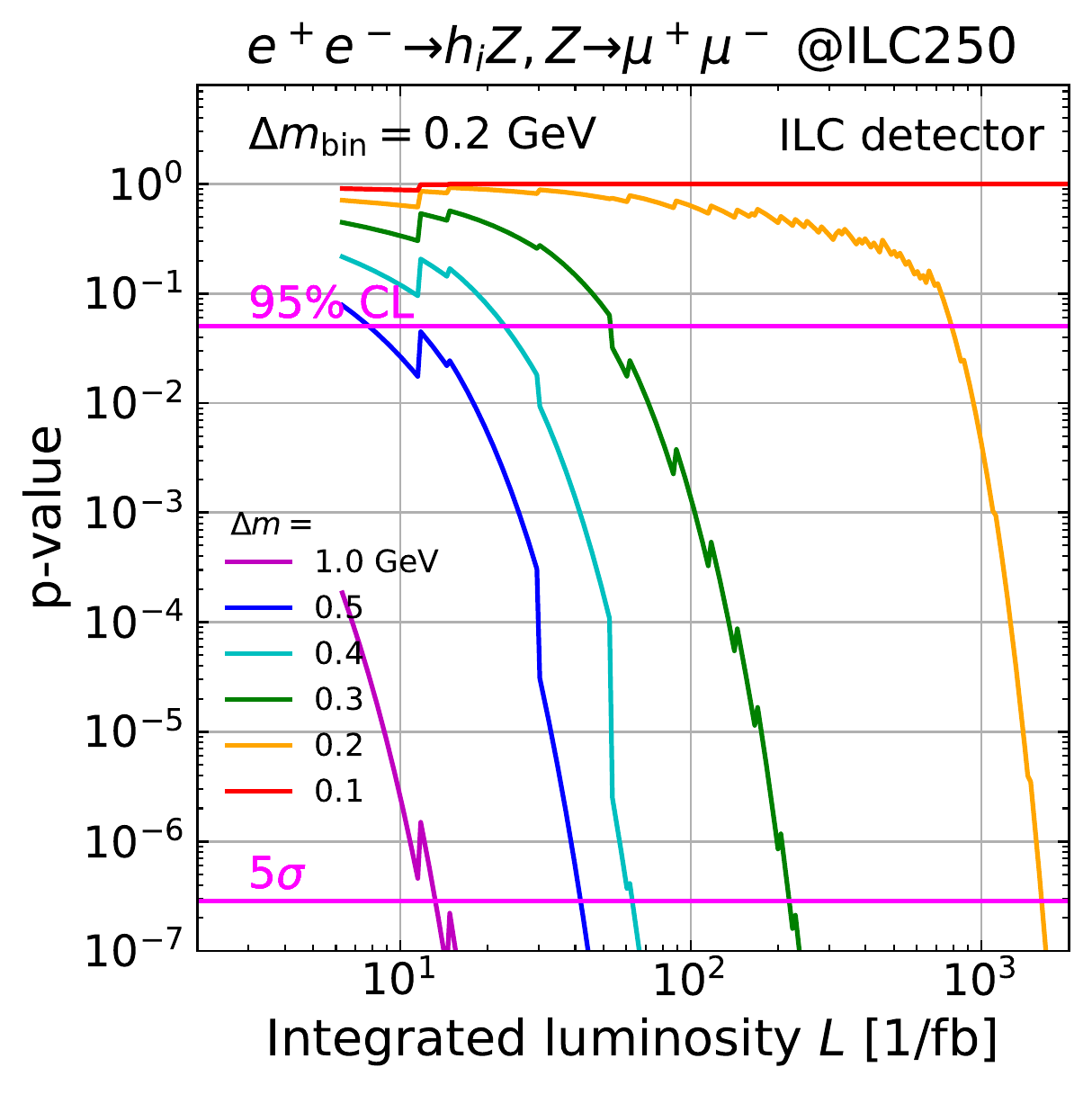}
 \includegraphics[width=0.325\textwidth]{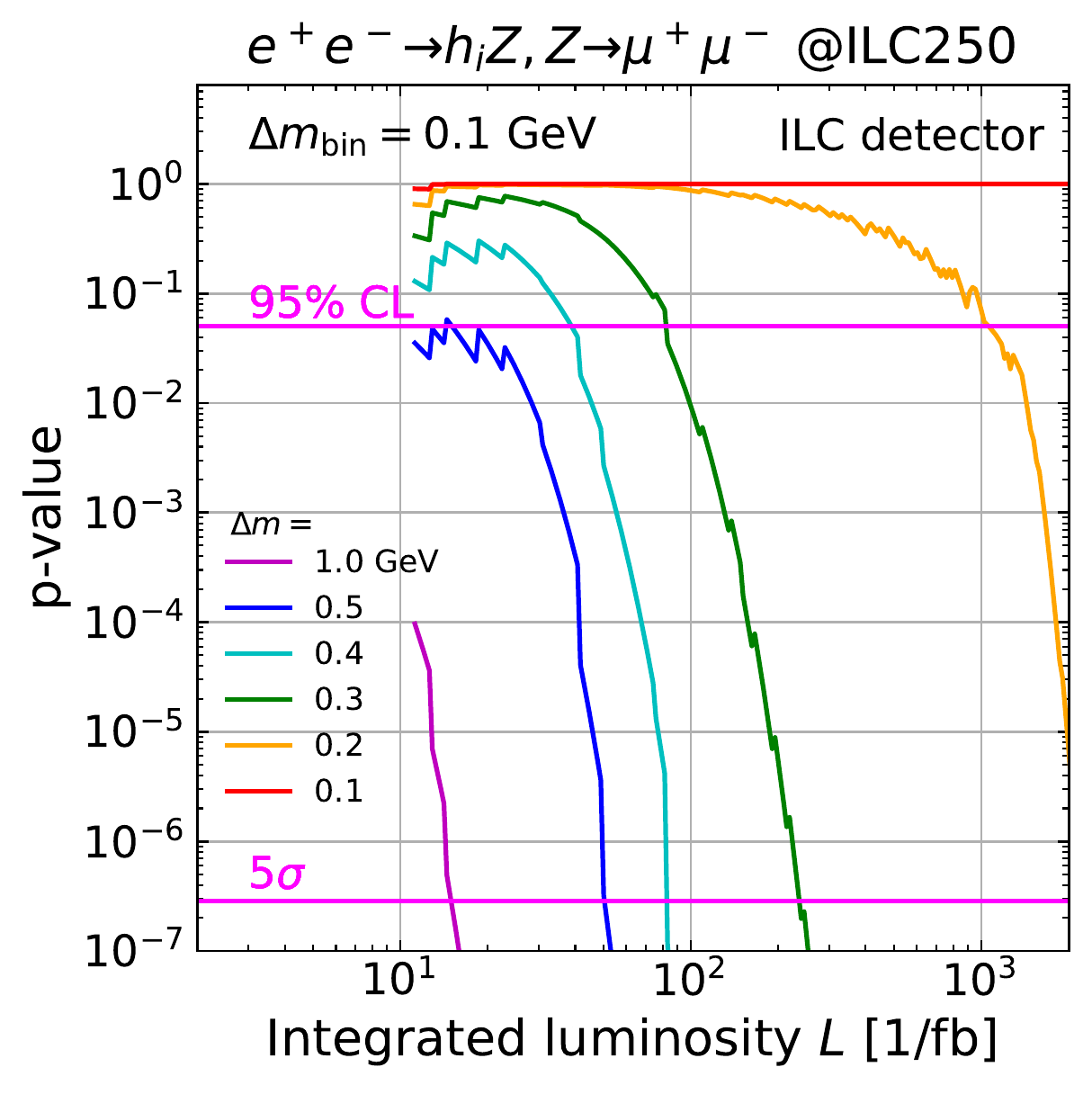}	 
 \caption{
  $p$-value for the distinction of the degenerate mass spectra
  as a function of the integrated luminosity for various $\Delta m$
  with the generic ILC detector.
  From the left to right panels, the size of bins in the 120--130~GeV range is taken
  as 0.4, 0.2, and 0.1~GeV, respectively.
 }
 \label{fig:pval}
\end{figure} 

Figure~\ref{fig:pval} (left) shows $p$-value calculated from the $\chi^2$
for the distinction of the degenerate mass spectra of various $\Delta m$
as a function of the integrated luminosity $L$ up to 2 ab$^{-1}$ planned at the ILC250,
where we take $N=25$ bins in the 120--130~GeV range, i.e. $\Delta m_{\rm bin}=0.4$~GeV. 
We find that, e.g.
the scenario with $\Delta m<0.3$~GeV is excluded (discovered) with $L\sim30$~\ifb
(240~\ifb), while that with $\Delta m<0.2$~GeV is $L\sim520$~\ifb (1300~\ifb).

We also investigate the effect of the bin size on the distinction; 
from the left to right panels, we take the number of bins in the 120--130~GeV range 
as $N=25$, 50 and 100, 
i.e. $\Delta m_{\rm bin}=0.4,\ 0.2,\ 0.1$~GeV, respectively.
Even for $\Delta m_{\rm bin}>\Delta m$, we expect a sensitivity to exclude or discover 
the degenerate-scalar scenario  
within the planned integrated luminosity 2 ab$^{-1}$ if $\Delta m>0.2$~GeV.
A better sensitivity can be expected with more optimized analyses,
and more dedicated studies such as including the background and the systematic errors should be reported elsewhere.

\section{Summary}\label{summary}

In this paper, we have studied a pseudoscalar DM model arising from the CxSM.
The scalar potential we adopted in our study is most general and renormalizable,
in which the global symmetry is softly broken by the operators up to the mass dimension two
so that it is not suffered from the domain-wall problem
as different from the minimal pseudo Nambu--Goldstone DM model. 
We showed that the DM--nucleon scattering amplitudes 
mediated by two scalar particles $h_1$ and $h_2$ are cancelled
when the masses of two scalars are degenerate,
and investigated the allowed model parameter space of the degenerate scenario
under the direct detection experiments and the measurements of the relic density of the DM. 

In addition, we discussed a possibility to verify such a degenerate-scalar scenario
by using the recoil mass technique at the ILC. 
The recoil mass is independent of how the degenerate scalars decay,
and hence the analysis is independent of the mass hierarchy
between the DM and the degenerate scalars in the model.
We found that a pair of states separated by 0.2~GeV can be distinguished
from the single SM-like Higgs state
at 5$\sigma$ with integrated luminosity of 2~ab$^{-1}$.

Finally, we briefly mention constraints on the degenerate scalar scenario from so-called indirect detection experiments, which restrict the annihilation processes of the DM. 
We expect that the degenerate scalars suppress the annihilation rates of the DM pair to fermions or gauge bosons owing to the cancellation mechanism so that these processes are not severely constrained from the indirect detection experiments. This consequence is different from a typical Higgs portal DM model (e.g., \cite{Duerr:2015aka}), which does not have such a cancellation mechanism. 
In the degenerate scenario, as is mentioned in Sec.~\ref{supp}, the DM annihilation to scalar pairs, $\chi \chi \to h_i h_j$, is not suppressed.
Therefore, the indirect detection experiments might give some constraints
on the model parameter space, which will be reported elsewhere. 

\section*{Acknowledgments}
We would like to thank Federico Ambrogi, Chiara Arina, Jan Heisig and Olivier Mattelaer 
for their valuable help with {\sc MadDM}
and Takanori Kono for many helpful discussions.
We also thank Keisuke Fujii for useful information on the ILC. 
The work of G.C.C is supported in part by Grants-in-Aid for Scientific
Research from the Japan Society for the Promotion of Science (No.16K05314).
The work of K.M. is supported in part by JSPS KAKENHI Grant No. 18K03648, 20H05239 and 21H01077.


\bibliography{bibcxsm}
\bibliographystyle{JHEP}

\end{document}